\documentclass[aps,amsmath,amssymb,superscriptaddress,showpacs,nofootinbib,10pt]{revtex4-1}
\usepackage{amsmath,amssymb}
\usepackage{slashed}
\usepackage{graphicx}
\usepackage{subfigure}
\usepackage{url}
\usepackage{bigints}
\newcommand{\beq}{\begin{equation}}
\newcommand{\eeq}{\end{equation}}

\newcommand{\be}{\begin{eqnarray}}
\newcommand{\ee}{\end{eqnarray}}
\long\def\hidestart#1\hideend{}
\begin{document}
\title{Effects of boundary conditions and gradient flow
in 1+1 dimensional lattice $\phi^4$ theory  
}
\author{A. Harindranath}
\email{a.harindranath@saha.ac.in}
\affiliation{Theory Division, Saha Institute of Nuclear Physics \\
 1/AF Bidhan Nagar, Kolkata 700064, India}
\author{Jyotirmoy Maiti}
\email{jyotirmoy.maiti@gmail.com}
\affiliation{Department of Physics, Tehatta Government College,\\
Tehatta, Nadia, West Bengal 741160, India}
\date{March 18, 2017}
\begin{abstract}
In this work we study the effects of boundary condition and gradient flow in 1+1
dimensional lattice $\phi^4$ theory. Simulations are performed with periodic (PBC)
and open (OPEN) boundary conditions in the temporal direction and the lattice fields
are then smoothed by applying gradient flow. Our results with observables such as
the $\langle\mid\hspace{-0.2em}\phi\hspace{-0.2em}\mid\rangle$ and the susceptibility
indicate that at a given volume, the phase transition point is shifted towards a lower
value of lattice coupling $\lambda_0$ for fixed $m_0^2$ in the case of OPEN 
as compared to the 
PBC, with this shift found to be diminishing as the volume increases. We have
employed the finite size scaling (FSS) analysis to obtain the true critical
behavior, mainly to emphasize the necessity of an FSS formalism incorporating the 
surface effect in the case of the open boundary. Above features have been found to be illuminated more clearly
by the application of gradient flow. Finally we compare and contrast the extraction of
the boson mass from the two point function (PBC) and the one point function (OPEN) as
the coupling, starting from moderate values, approaches the critical value corresponding
to the vanishing of the mass gap. In the critical region, finite volume effects become
dominant in the latter. The surface effect seems to be resulted in a less sharper phase
transition for OPEN compared to the PBC for all observables studied here.
\end{abstract}

\maketitle

\section{Introduction and Motivation}

The quantum field theory of 1+1 dimensional $\phi^4$ interaction, in spite of
its apparent simplicity, has a very rich structure and hence has been the
testing ground for various new non-perturbative approaches towards the field
theory. It also provides ample opportunity to study newly proposed algorithms
and calculational techniques. Extensive numerical results have been presented
in this theory in an earlier work \cite {De:2005ny}. In this work we present
numerical studies in this theory in the context of comparison between periodic
(PBC) and open boundary conditions (OPEN) in the temporal direction and the
effects of gradient flow (also known as Wilson flow).

In addition to the periodic boundary conditions in both the temporal and 
spatial
directions, one can have other types of boundary conditions. For the scalar
field, for example, one can have anti-periodic boundary condition in the
spatial direction (APBC). In the latter case one can study quantum kinks
\cite{De:2005}. Lattice Quantum ChromoDynamics (LQCD) conventionally uses
PBC in both the temporal and spatial directions for the gauge field. However,
in this case, the spanning of gauge configurations over different topological
sectors becomes more and more difficult as the continuum limit is approached.
As a remedy, open boundary condition in the temporal direction has been
proposed \cite{open0, open1, open2}. Numerical studies in pure Yang-Mills
theory \cite{opentopo,Chowdhury:2014kfa,Chowdhury:2014mra,Chowdhury:2015hta,
Amato:2015ipe} and QCD \cite{Bruno:2014jqa} have yielded encouraging results.

In order for OPEN to be effective, boundary artifacts should be negligible so
that one has a bulk region of considerable extent. This is possible as long
the system is not gap less (critical) \cite{shibata}. The success of OPEN in 
pure Yang-Mills theory and QCD hinges on the existence of a mass gap in these
systems, namely glueball and pion respectively. On the other hand, 1+1
dimensional lattice $\phi^4$ theory offers an opportunity to investigate in
detail the artifacts induced by the open boundary condition as the system
approaches criticality.   

For extracting various observables in lattice field theories, smoothing of
lattice fields is essential, in order to overcome lattice artifacts. The
gradient (Wilson) flow \cite{wf1, wf2, wf3} provides a very convenient tool
for smoothing, with a rigorous mathematical underpinning. For some recent
studies of gradient flow in the context of scalar theory, see Refs. 
\cite{mo, fujikawa, dgk, monahan}. In our previous work on Yang-Mills
theory, we have demonstrated the effectiveness of gradient flow in the
extraction of the topological susceptibility and 
glueball masses. Thus it will be very interesting
to study the effect of gradient flow on various observables, independent of
the boundary condition,
 in 1+1 dimensional lattice $\phi^4$ theory.        
 
\section{Boundary Conditions and Gradient Flow}
In the continuum, the Euclidean Lagrangian (density) for $\phi^4$ theory is
given by 
\be
{\mathcal{L}} = \frac{1}{2} \partial_\mu \phi \partial_\mu \phi 
~+~ \frac{1}{2} m^2 \phi^2 ~+ \frac{\lambda}{4!}\phi^4.
\ee

On a periodic (on all space-time directions) lattice, the Euclidean action in
`d' space-time dimension is conventionally written as
\cite{De:2005ny}
\be
S = -\sum_x \sum_\mu \tilde\phi_x \tilde\phi_{x+\mu} +
\left(d+\frac{m_0^2}{2}\right)\sum_x \tilde\phi_x^2 + \frac{\lambda_0}{4!}\sum_x
\tilde\phi_x^4
\ee
where all the parameters and fields have been made dimensionless by
multiplying them with appropriate powers of lattice spacing `$a$'.

However, in order to impose open boundary condition in the temporal
direction we follow the construction of the transfer matrix. For this purpose 
the
lattice action is written in terms of time slice action density $E(t)$ as
$S = \sum\limits_{t} E(t)$ and the kinetic term in the temporal direction is
distributed symmetrically around the time slice `$t$'. The details are  given
below.

First, with the aid of forward and backward lattice derivatives
\be
\partial^f_\mu \phi = \frac{1}{a}\left( \phi_{x+\mu} -\phi_x \right)~~~
{\rm and}
~~~ \partial^b_\mu \phi = \frac{1}{a}\left( \phi_{x} -\phi_{x-\mu} \right)~,
\ee
we write the symmetrized expression for kinetic term as 
\be
\frac{1}{2}\partial_\mu \phi \partial_\mu \phi &=& \frac{1}{4} 
\left( \partial^f_\mu \phi \partial^f_\mu \phi + \partial^b_\mu \phi
\partial^b_\mu \phi \right) \nonumber \\
&=& \frac{1}{4a^2} \left(~ 2 \sum_\mu \phi_x^2 ~+~ \phi^2_{x+\mu}
~+~ \phi^2_{x-\mu} ~-~2 \phi_x \phi_{x+\mu} ~-~2 \phi_x \phi_{x-\mu}
\right)~.
\ee


This enables one to write down the time slice action density for periodic
lattice as
\be
E_{\rm PBC}(t)= {\cal{T}}_{t}~+~{\cal{V}}_{t}~+~ \frac{1}{2} {\cal{T}}_{t,t+1} ~+~
\frac{1}{2} {\cal{T}}_{t-1,t}  
\ee  
where
\be
{\cal{V}}_t &=& \sum_{\vec{x}} ~\left(~ \frac{1}{2} a^2m^2\phi^2_{\vec{x},t}~+~ 
\frac{a^2\lambda}{4!} \phi^4_{\vec{x},t}~\right)~, \nonumber\\
{\cal{T}}_t  ~&=&~  \frac{1}{4}\sum_{\vec{x},k} \Big ( 2~\phi^2_{\vec{x},t} 
~+~ \phi^2_{\vec{x}+\hat{k},t} ~+~ \phi^2_{\vec{x}-\hat{k},t} ~-~2 ~\phi_{\vec{x},t}
~\phi_{\vec{x}+\hat{k},t} ~-~2 ~\phi_{\vec{x},t}
~\phi_{\vec{x}-\hat{k},t} \Big)~,\nonumber
\\
{\cal{T}}_{t,t+1} & = & \frac{1}{2} \sum_{\vec{x}} \Big ( \phi^2_{\vec{x},t}
~+~ \phi^2_{\vec{x},t+1} ~-~2 ~\phi_{\vec{x},t} ~\phi_{\vec{x},t+1} \Big)~,\nonumber \\
{\cal{T}}_{t-1,t} & = & \frac{1}{2} \sum_{\vec{x}} \Big ( \phi^2_{\vec{x},t}
~+~ \phi^2_{\vec{x},t-1} ~-~2 ~\phi_{\vec{x},t} ~\phi_{\vec{x},t-1} \Big)~ \nonumber
\ee
with $\hat{k}$ being the unit vector in an arbitrary spatial direction.



Following this definition, we denote $e^{-H_m\left(t,t+1\right)}$ as the general
transfer matrix element between the time slices $t$ and $t+1$ where
\be
H_m\left(t,t+1\right) = \frac{1}{2} \left({\cal{T}}_t+{\cal{V}}_t\right) ~+~
\frac{1}{2} \left({\cal{T}}_{t+1} + {\cal{V}}_{t+1}\right)~+~ {\cal{T}}_{t,t+1}.
\ee

Particularly for a lattice of temporal extent `$T$', the transfer matrix element
between the boundary time slices $t=T-1$ and $t=0$ is determined by
\be
H_m\left(T-1,0\right) ~&=&~ \frac{1}{2} \left({\cal{T}}_{T-1}+{\cal{V}}_{T-1}
\right) ~+~ \frac{1}{2} \left({\cal{T}}_{0} + {\cal{V}}_{0}\right)~+~
      {\cal{T}}_{T-1,0}\nonumber
\ee


Now, if the temporal boundary becomes open, the corresponding term drops out 
from
the partition function. This leads us to relate the actions for lattices with
two different boundary conditions in the temporal 
direction (periodic and open) as
$S_{\rm PBC} = S_{\rm OPEN} + \Delta S$ where $\Delta S = H_m\left(T-1,0\right)$.

The absence of the term $\Delta S$ from the action for a lattice with open
boundary (temporal direction), in turn, also modifies the expressions for action
densities at the temporal boundaries. They are given by
\be
E_{\rm OPEN}(t=0) &=& \frac{1}{2}\left({\cal{T}}_{0}~+~{\cal{V}}_{0}\right)~+~
\frac{1}{2} {\cal{T}}_{0,1}\\
     {\rm and}\quad E_{\rm OPEN}(t=T-1) &=& \frac{1}{2}\left({\cal{T}}_{T-1}~+~
     {\cal{V}}_{T-1}\right)~+~\frac{1}{2} {\cal{T}}_{T-2,T-1}.
\ee
Within the bulk ($0<t<T-1$), $ E_{\rm OPEN}(t)~=~ E_{\rm PBC}(t)$.



In order to smooth the lattice $\phi$ field, gradient flow is used. 
For $\phi^4$ theory it is known that, in four dimensions, there are 
potential
divergences in the correlation functions for flow time greater than zero and
for this reason in Refs. \cite{mo} and \cite{dgk} simple flow equation
corresponding to free field theory has been used. Fujikawa \cite{fujikawa}
has proposed a modification of the flow equation to tackle this problem. 
However, the $\phi^4$ in 1+1 dimensions which we study here is free from
such divergences. 
Here we follow the choice made in our previous works for Yang-Mills theory, 
namely, picking
the gradient of the action to drive the flow. Thus   
for $\phi^4$ theory in the Euclidean space in 1+1 dimensions, in the 
continuum,
 the flow equation is chosen to be
\be
\frac{\partial \psi\left(x,\tau\right)}{\partial \tau} &=& -\frac{\delta S
\left[\psi\right]}{\delta \psi\left(x,\tau\right)}\nonumber\\
&=& \partial_\mu \partial_\mu
\psi(x,\tau) - m^2  \psi(x,\tau)~ - \frac{\lambda}{6} \psi^3(x,\tau)
\ee 
where $\psi(x,\tau=0) = \phi(x)$ with `$\tau$' being the flow time.

We numerically solve this equation on the lattice using the 2$^{\rm nd}$
order Runge-Kutta method.
\section{Extraction of boson mass from lattices with different boundaries}
For the extraction of boson mass we have used the simplest and the most familiar
scalar operator - the time sliced field $\phi(t)=\frac{1}{V}\sum\limits_{\vec{x}}
\phi\left(\vec{x},t\right)$ where $V$ is the spatial volume of the lattice. The
mass can be easily extracted from the two point correlation function for this
scalar operator which, in the case of periodic boundary in the temporal direction,
behaves as
\be 
G(t) =  \left\langle \phi(t) \phi(t=0)  \right\rangle_{\rm
PBC}
&\approx&   C_0~+~ C_1 ~ \Big [ 
e^{-mt}~ + ~ e^{-m(T-t)}
\Big ]~\nonumber\\
&=& C_0 ~+~ 2 C_1 ~ e^{-m T/2}~\cosh m\Big(\frac{T}{2}-t\Big) \label{ansatz}
\ee 
where
\be
C_1 = \frac{\left|\langle
0\left|\phi\left(0\right)\right|B\rangle\right|^2}{2m} 
\ee
with $\mid B\rangle $ being the one boson state. To improve statistics, one can
average over the source time as well. The effective mass can be evaluated by
solving the equation $F(m)=0$
\be
{\rm where}\quad F(m)~&=&~ (r_1-1)~\Big [ \cosh m(\Delta t-1) ~ - ~   \cosh m \Delta t \Big ]
~+~ (1-r_2)~ \Big [ \cosh m(\Delta t+1) ~ - ~   \cosh m \Delta t \Big ]
\label{meff1}
\\
{\rm with}\quad 
r_1  ~&=&~ \frac{G(t-1)}{G(t)}, ~~~~ 
r_2  ~=~ \frac{G(t+1)}{G(t)}~~~~~ {\rm and}~~ \Delta t=T/2 - t~.
\label{meff2}
\ee

Note that the boson mass extracted from the propagator 
using Eq. (\ref{ansatz}) is the pole mass 
and hence is the physical
mass \cite{mm}, independent of the lattice spacing.  

In the case of open boundary in the time direction, to avoid the boundary effects 
one
needs to be well within the bulk while computing the two point correlation
function. However, the second exponential will be absent from the expression of
the two point function due to the loss of periodicity. On the other hand, as
the time translational invariance is also lost in this case, one cannot average
over the source time as well. However, within the bulk, well away from the
boundary region translational invariance is recovered. So one can take average
over few time slices to regain statistics.

However, this effort breaks down as one approaches the critical region. It will
be shown later on in this study that, the effect of open boundary starts to
engulf the whole bulk region as we move towards the critical point. Mass
extraction from two point function becomes almost impossible.

Surprisingly, the open boundary itself opens up new pathways to extract the
mass. Following Ref. \cite{Chowdhury:2015hta}, in this section we review how
the boson mass can be extracted from a one-point function in the case of
open boundary using
a generic time sliced scalar operator ${\cal O}(t)$.
 
We start from
\be
\left\langle {\cal O}(t) \right\rangle_{\rm OPEN}
&=& ~
\frac{\bigintss {\mathcal D}\phi~{\cal O}(t)~e^{-S_{\rm OPEN}}} 
{\bigintss{\cal D}\phi  ~e^{-S_{\rm OPEN}} } \nonumber \\
&=&{\bigintss {\cal D}\phi~  {\cal O}(t)~ e^{-S_{\rm PBC}+\Delta S }\Big/ 
\bigintss {\cal D}\phi~ e^{-S_{\rm PBC}}
\over
\bigintss {\cal D}\phi~ e^{-S_{\rm PBC}+\Delta S } \Big/ 
\bigintss {\cal D}\phi~ e^{-S_{\rm PBC}} }\nonumber\\
&=& \left\langle {\cal O}(t)\right\rangle_{\rm PBC} 
~+~\frac{\left\langle{\cal O}(t)~e^{\Delta S }  \right\rangle_{\rm
PBC}^{\rm connected}}{\left\langle e^{\Delta S}
\right\rangle_{\rm PBC}}~\label{mastere} \\
 &=& \left\langle{\cal O} (t)\right\rangle_{\rm PBC} ~+~ \frac{1}{r}
\left\langle {\cal O}(t)~e^{H_m(T-1,0)}\right\rangle_{\rm PBC}^{\rm
connected}
 \label{oneptwop}
\ee
where $r= \langle e^{\Delta S}\rangle_{\rm PBC} = \langle e^{H_m(T-1,0)}
\rangle_{\rm PBC}$.
As $e^{H_m(t,t+1)}$ is also a scalar operator, from Eq. (\ref{oneptwop}) 
we have
\be
\left\langle {\cal O}(t)\right\rangle_{\rm OPEN} 
\approx  \left\langle {\cal O}(t)\right\rangle_{\rm PBC}
~+~2C_{1}^{\prime} ~e^{-m T/2}~\cosh m\Big(\frac{T}{2}-t\Big).
\label{onepmass}
\ee
where $m$ is the scalar boson mass.

Thus we find that one can extract certain two-point correlators computed
with periodic boundary condition in the temporal direction by analyzing
the data for the functional average of a scalar operator (one-point function) 
computed with open boundary (in the temporal direction) in the region of $t$
where it differs, due to the breaking of translational invariance, from the
same computed with periodic boundary. Now due to time translational invariance
in the case of PBC, we can assume $\left\langle {\cal O}(t)\right
\rangle_{\rm PBC}$ to be constant and the evaluation of effective mass can
then be done again following the Eqs. (\ref{meff1}) and (\ref{meff2}).

\section{Simulation details}
As we have restricted ourselves to $1+1$ dimensions within this study, the
fields are dimensionless here. The notations for the bare parameters of the
theory on the lattice are chosen to be $m_0^2 = a^2m^2$ and $\lambda_0 =
a^2\lambda$. As we know, for the stability of the theory one must have
$\lambda_0 \ge 0$ and the phase transition associated with the spontaneous
breaking (or restoration) of $Z(2)$ symmetry takes place only for $m_0^2 < 0$.

For the study with periodic boundary in both directions, following the method
of Brower and Tamaya \cite{bt}, we have used Wolff's single cluster algorithm
\cite{wolff2,wolff} blended with the standard metropolis algorithm in 1:1
ratio for the
generation of field configurations. For the details of the whole procedure, see
\cite{De:2005ny}. However, in the study with open boundary in the temporal direction
(periodic in spatial direction), we resort only to the standard metropolis
algorithm for configuration generation.

Following the discussions in \cite{De:2005ny}, 
we have used $\langle\mid\hspace{-0.2em}\phi\hspace{-0.2em}\mid\rangle$ as 
the order parameter to investigate the phase structure. Here, 
$\phi = \sum\limits_{{\rm sites}~x}\phi\left(x\right)/{\rm lattice~volume}$. 
The absolute value is taken to avoid the effect of tunnelling enforced by the 
algorithm of configuration generation. For the phase diagram we mainly 
resorted to the former study \cite{De:2005ny}. Here, we will study the 
effects of gradient flow and boundary conditions on the phase diagram in due 
course.

In this work, we explored the phase structure and the spectrum of
the theory for two different sets of bare parameters given by $m_0^2 = -0.5$
and $m_0^2 = -1.0$. For each set of parameters (i.e., $m_0^2$ and $\lambda_0$),
we have first discarded $10^6$ configurations for thermalization and then
generated another $10^8$ configurations for the measurement purpose. Measurements
are done on one in every thousand configurations. In order to study the finite size
effects on the results, whole investigations have been done with four different
lattice volumes such as $48^2$, $64^2$, $96^2$ and $128^2$.

Configurations chosen for measurements are placed under gradient flow which 
is run for
one hundred steps, to be called as flow level, with a 
stepsize of $\delta \tau = 0.02$. Measurements are done after every ten flow 
levels in addition to the measurement done
before the flow is started.

\section{Numerical results}
In this section we present and discuss our numerical results showing the effects of the boundary conditions and the gradient flow on various observables of interest. In principle, one could divide this into two separate studies altogether as the gradient flow and the boundary condition are two completely disjoint categories. 
Neither of them has anything to do with the other. However, on one hand, we wanted to study the 
smoothening utility of the gradient flow irrespective of the choice of the boundary condition and on the other hand we have a different aim to study the effect of boundary condition as well. Thus they have come together in the same discussion here. Nevertheless they are independent of each other.

\subsection{The field variable}
As the translational invariance is lost in the temporal direction when open boundary condition is imposed in that direction, time sliced scalar field $\phi(t)$ could serve as an important observable to study the effect of open boundary as compared to periodic boundary. For the reasons stated earlier, here too, we take absolute value before evaluating the configuration average.

\begin{figure}
\begin{center}    
\includegraphics[width=0.5\textwidth]{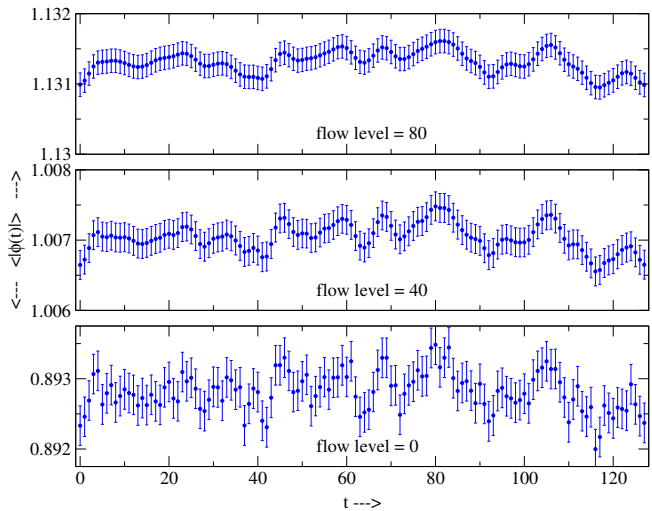}
\caption{Plot of expectation value of $|\phi(t)|$ for $m_0^2=-0.5$, $\lambda_0 =1.65$ and
$L=128$ with PBC for three different gradient flow levels.}
\label{tsppbc}
\end{center}
\end{figure}

\begin{figure}
\begin{center}   
\includegraphics[width=0.5\textwidth]{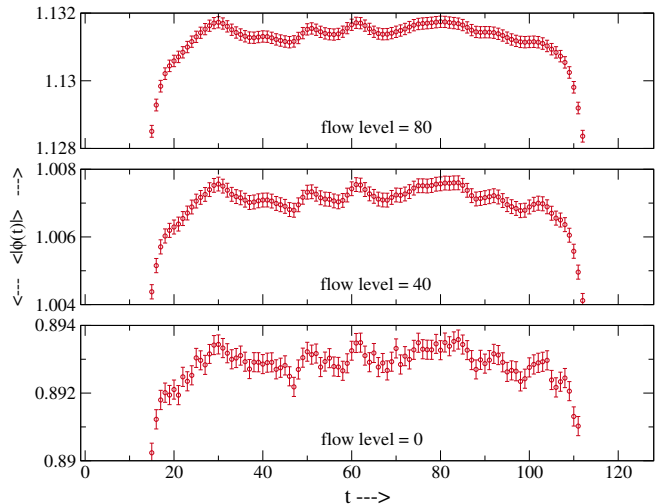}
\caption{Plot of expectation value of $|\phi(t)|$ for $m_0^2=-0.5$, $\lambda_0 =1.65$ and
$L=128$ with OPEN for three different levels of gradient flow.}
\label{tspopen}
\end{center}
\end{figure}

In Figs. \ref{tsppbc} and \ref{tspopen}, respectively for periodic and open 
boundary conditions in temporal directions, we present the expectation value of $|\phi(t)|$ in 
three different subdiagrams - one without any gradient flow and two others 
with two different levels of gradient flow all for a particular set of lattice 
parameters $m_0^2 = 0.5$, $\lambda_0 = 1.65$ and $128^2$ lattice. The figures 
clearly show the smoothening effect with the increase of flow level. Note 
that, values of $\langle\mid\hspace{-0.2em}\phi(t)\hspace{-0.2em}\mid\rangle$ 
are gradually rising up with increasing flow level. The widths of windows for 
the values of $\langle|\phi(t)|\rangle$ are taken to be same in all the 
three subdiagrams instead of taking them to be proportional to the respective 
average values. This actually has reduced the manifestation of smoothening 
effect to some extent.

\begin{figure}
\begin{center}   
\includegraphics[width=0.5\textwidth]{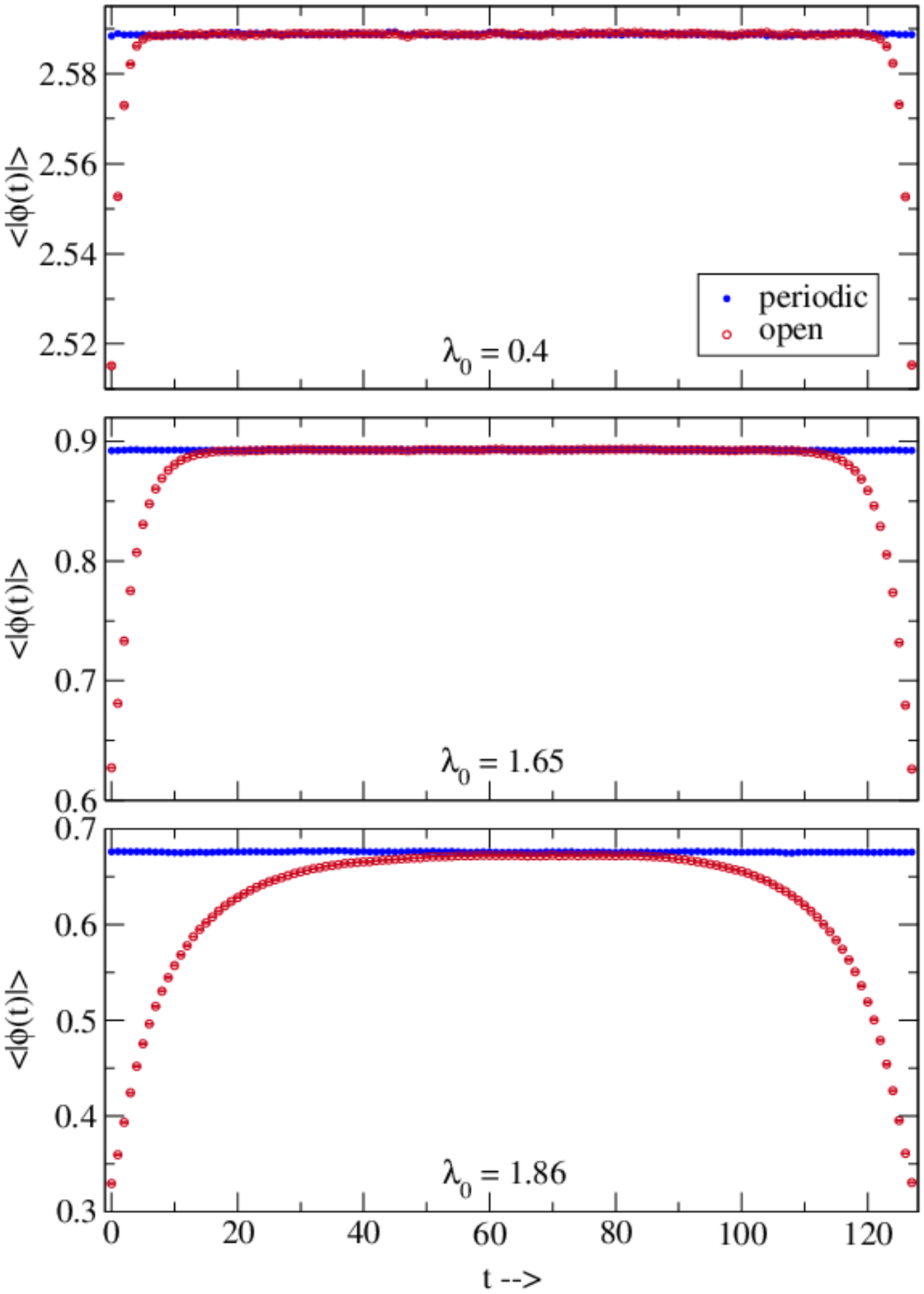}
\caption{Comparison of $\langle|\phi(t)|\rangle$ between PBC and OPEN for $m_0^2=-0.5$,
$L=128$ and three different values of $\lambda_0$ without any gradient flow.}
\label{gfcompl0}
\end{center}
\end{figure}

\begin{figure}
\begin{center}   
\includegraphics[width=0.5\textwidth]{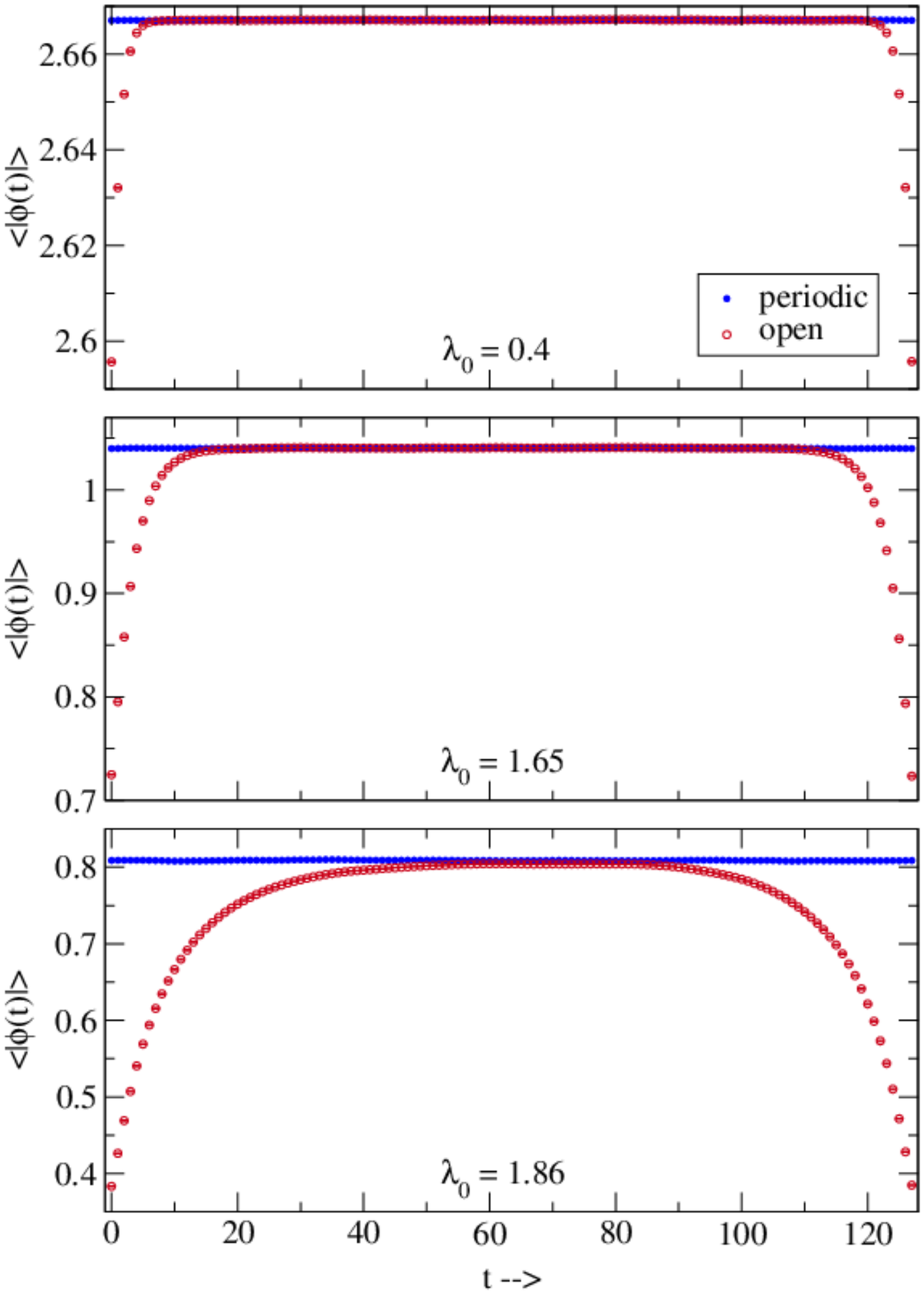}
\caption{Comparison of $\langle|\phi(t)|\rangle$ between PBC and OPEN for $m_0^2=-0.5$,
$L=128$ and three different values of $\lambda_0$ at gradient flow level 50.}
\label{gfcompl50}
\end{center}
\end{figure}

To make exhibition of boundary effect clearer, in Fig. \ref{gfcompl0}, we compare
the behaviour of $\langle\mid\hspace{-0.2em}\phi(t)\hspace{-0.2em}\mid\rangle$
for  $m_0^2=-0.5$ and $L=128$ for periodic and open
boundary conditions without any gradient flow for three different values of
$\lambda_0$ all in the broken symmetric phase. For the smallest value of the
coupling which is far away from the critical point, effects of open boundary
are found to be only at the edges leaving a long bulk region
matching with the counterpart in PBC. As the coupling increases one gets closer
to the region of phase transition and effect of open boundary extends on both
side squeezing the bulk. For $\lambda_0 = 1.86$, we are already in the critical
region and we observe that the bulk region has almost vanished. The same is
presented in Fig. \ref{gfcompl50} for gradient flow level 50. We 
notice that it
only smoothens the data (which is, although, hardly visible here because of 
the
wide scale of values 
for $\langle\mid\hspace{-0.2em}\phi(t)\hspace{-0.2em}\mid
\rangle$ covered in the figure) leaving the boundary effects unchanged.

\begin{figure}
\begin{center}   
\includegraphics[width=0.5\textwidth]{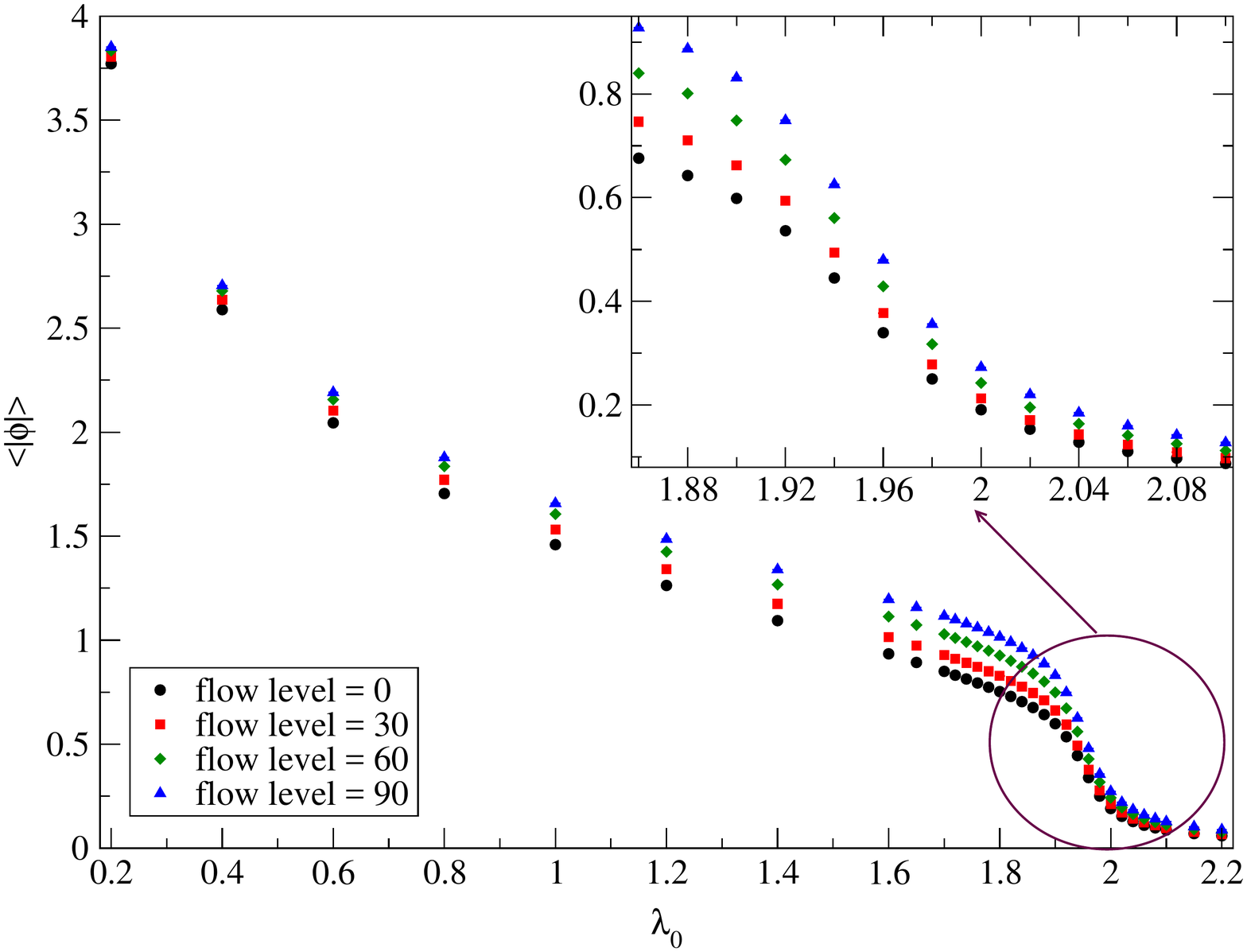}
\caption{Plot of $\langle\mid\hspace{-0.2em}\phi\hspace{-0.2em}\mid\rangle$
versus $\lambda_0$ for different values of the gradient flow level for
$m_0^2=-0.5$ and $L=128$ in the case of PBC.}
\label{phigf}
\end{center}
\end{figure}

In Fig. \ref{phigf} we plot $\langle \mid \hspace{-0.2em} \phi \hspace{-0.2em}
\mid \rangle$ versus $\lambda_0$ for different values of the gradient flow
level for $128^2$ lattice at $m_0^2 = -0.5$. The monotonous rise in the value of $\langle \mid \hspace{-0.2em} \phi \hspace{-0.2em}\mid \rangle$ with increasing
gradient flow level is consistent with the behaviour of $\langle \mid \hspace{-0.2em} \phi(t) \hspace{-0.2em}\mid \rangle$. Details of the behavior in the critical region are
shown in the inset of the figure. The trend is seemed to be retained across  
the phase transition with an additional feature exhibiting a sharper fall of the observable across the transition point as the flow level is increased.

\begin{figure}
\begin{center}   
\includegraphics[width=0.5\textwidth]{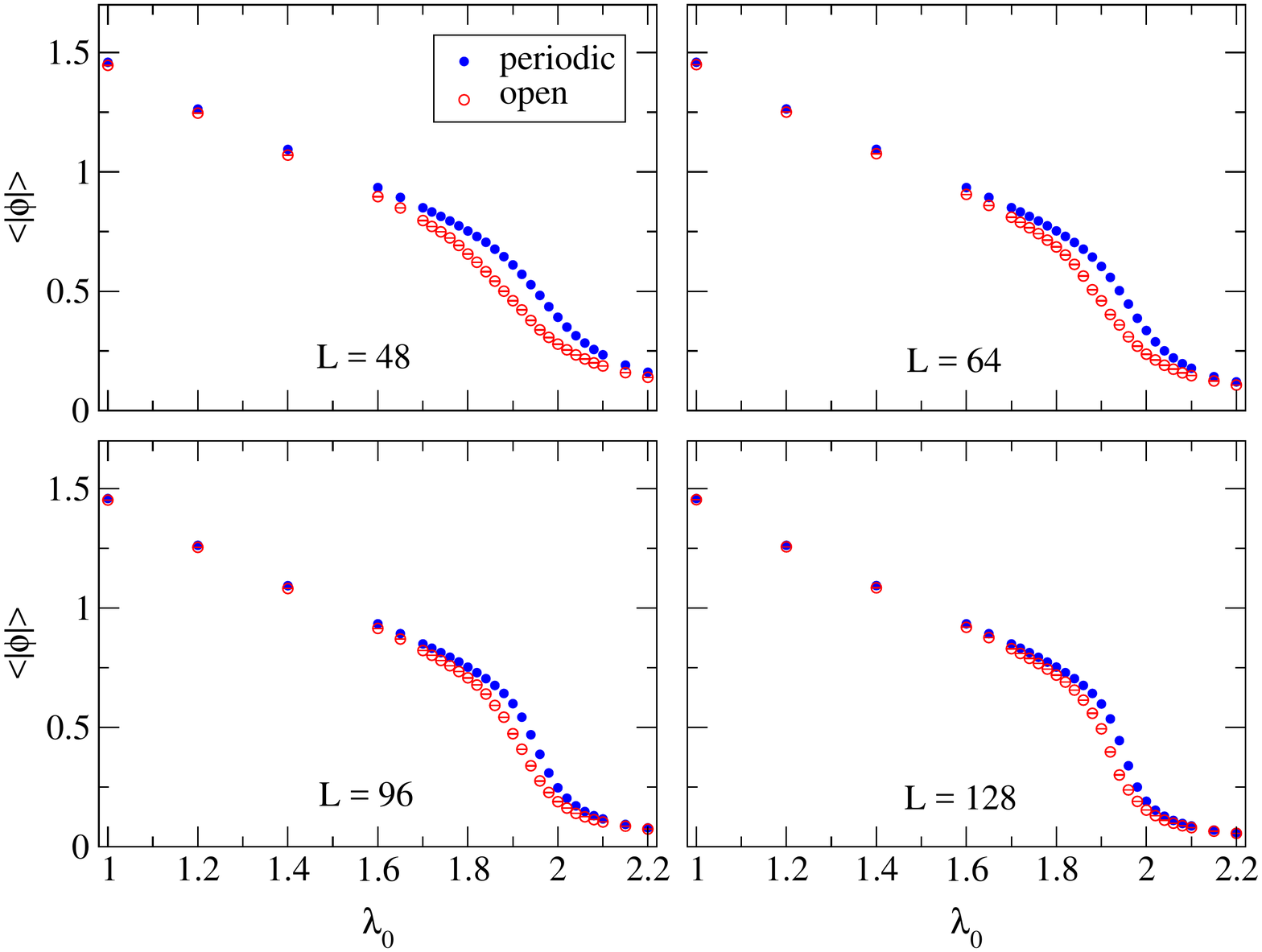}
\caption{Plot of $\langle\mid\hspace{-0.2em}\phi\hspace{-0.2em}\mid\rangle$
versus $\lambda_0$ for
PBC and OPEN without gradient flow for different $L$ at $m_0^2=-0.5$. }
\label{phi-po-gf0}
\end{center}
\end{figure}

\begin{figure}
\begin{center}   
\includegraphics[width=0.5\textwidth]{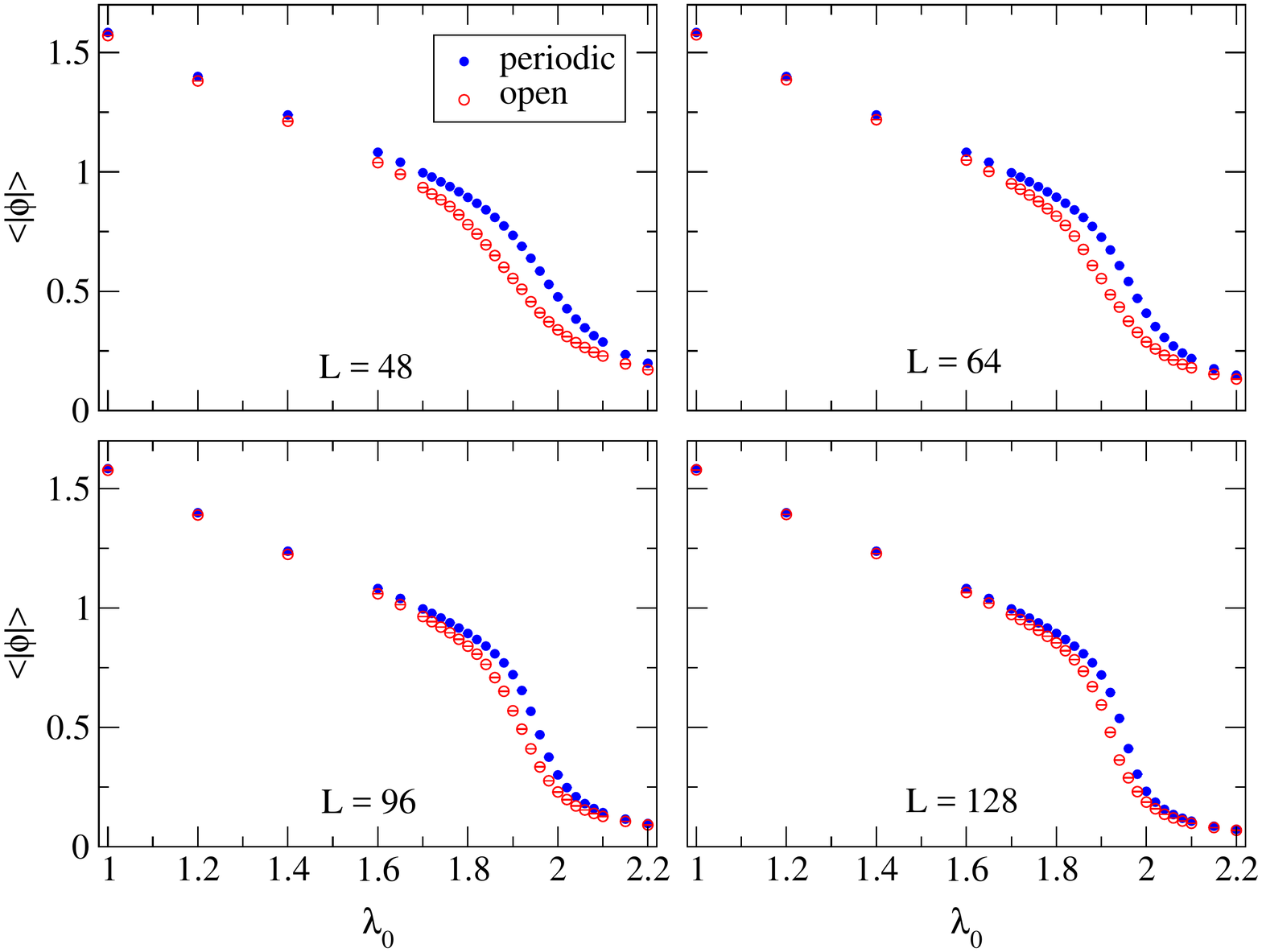}
\caption{Plot of $\langle\mid\hspace{-0.2em}\phi\hspace{-0.2em}\mid\rangle$
versus $\lambda_0$ for
PBC and OPEN at gradient flow level 50 for different $L$ at $m_0^2=-0.5$. }
\label{phi-po-gf50}
\end{center}
\end{figure}

In Fig. \ref{phi-po-gf0}, we present the comparison between $\langle \mid
\hspace{-0.2em} \phi \hspace{-0.2em} \mid \rangle$ computed using PBC and the
same obtained with OPEN for different lattice sizes at $m_0^2 = -0.5$ before
applying the gradient flow on the fields. Deep in both the broken-symmetric and
the symmetric phases, for all the lattice volumes, the results for PBC and OPEN 
are found to be matching within our statistical error. However, inside the critical
region a clear disagreement is observed. Phase transition appears to take place
at smaller values of $\lambda_0$ in the case of OPEN compared to the PBC. However,
the gap between the transition point in two different boundaries seems to be 
vanishing with the increase of lattice size. In addition, although not clearly
prominent, it appears that the fall of $\langle \mid\hspace{-0.2em} \phi
\hspace{-0.2em} \mid \rangle$ across the transition point is slightly sharper for
PBC in comparison with OPEN. Both these behaviors are consistent with
the expectation that the effect of boundary surface diminishes in infinite
volume limit. The picture remains to be almost unaltered by the application of
gradient flow other than raising the values universally little bit. This has
been emphasized in Fig. \ref{phi-po-gf50}.

\subsection{The susceptibility}
\begin{figure}
\begin{center}   
\includegraphics[width=0.5\textwidth]{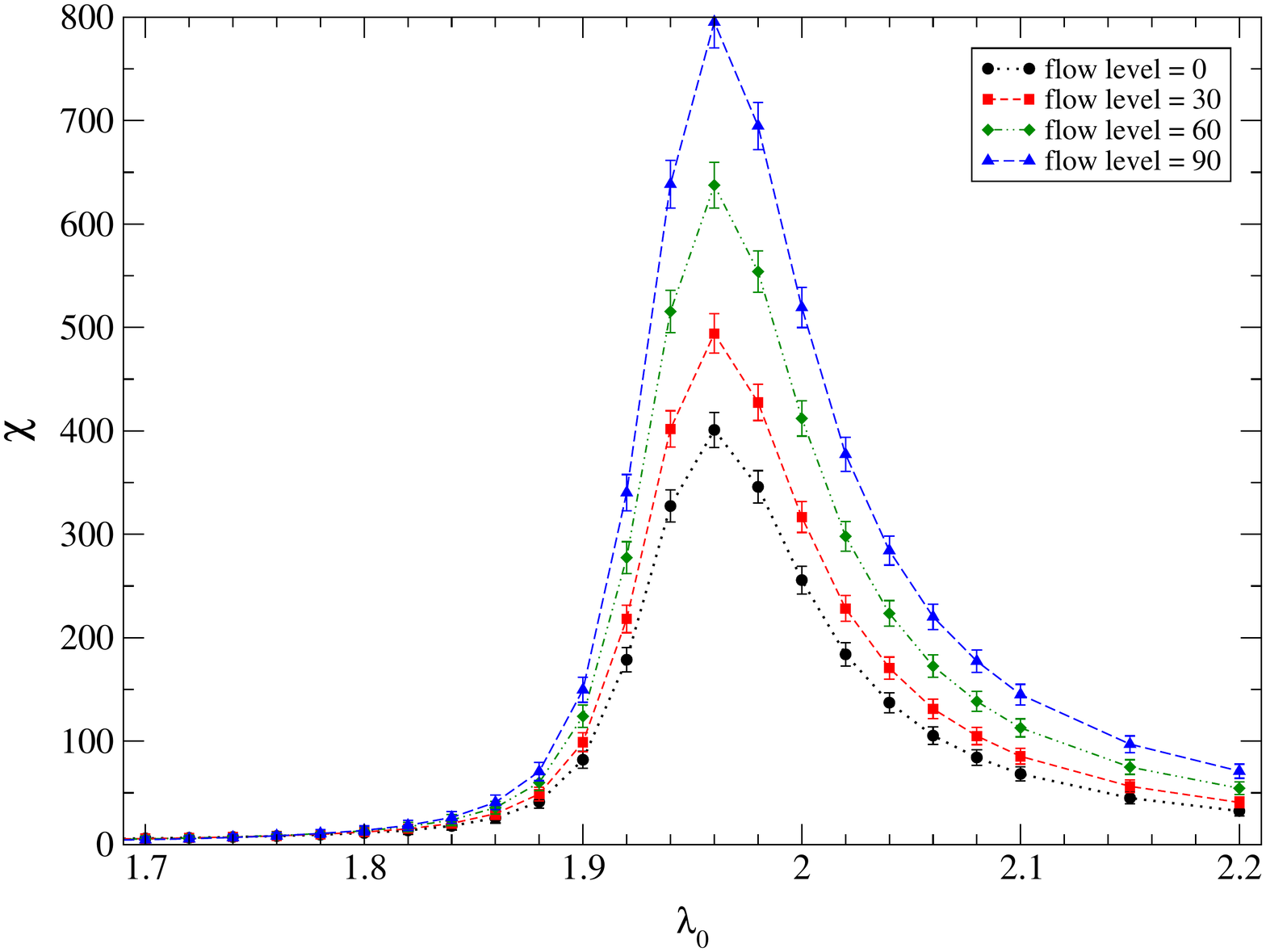}
\caption{Comparison of susceptibility for different levels of gradient flow
with PBC at $m_0^2=-0.5$ and $L=128$.}
\label{suscpbcwf}                                                               
\end{center}                                     
\end{figure}

In this subsection, we study the effects of gradient flow and the boundary conditions on the 
behaviour of another observable of interest, namely, the susceptibility defined as
$\chi  = \sum\limits_x \langle \phi(x) \phi(0) \rangle/{\rm lattice~volume}$.
In Fig. \ref{suscpbcwf}, we
show the behaviour of susceptibility as a function of $\lambda_0$ for
$m_0^2=-0.5$ with PBC on $128^2$ lattice for different levels of gradient flow.
Consistent with the trends as observed in the case of $\langle\mid\hspace{-0.2em}
\phi\hspace{-0.2em}\mid \rangle$, the peak of susceptibility becomes higher
and higher as the level of gradient flow increases leaving the peak position
unchanged. This behavior is expected since the gradient flow serves to
remove the lattice artifacts and brings the lattice theory closer to the
continuum limit.

\begin{figure}
\begin{center}   
\includegraphics[width=0.5\textwidth]{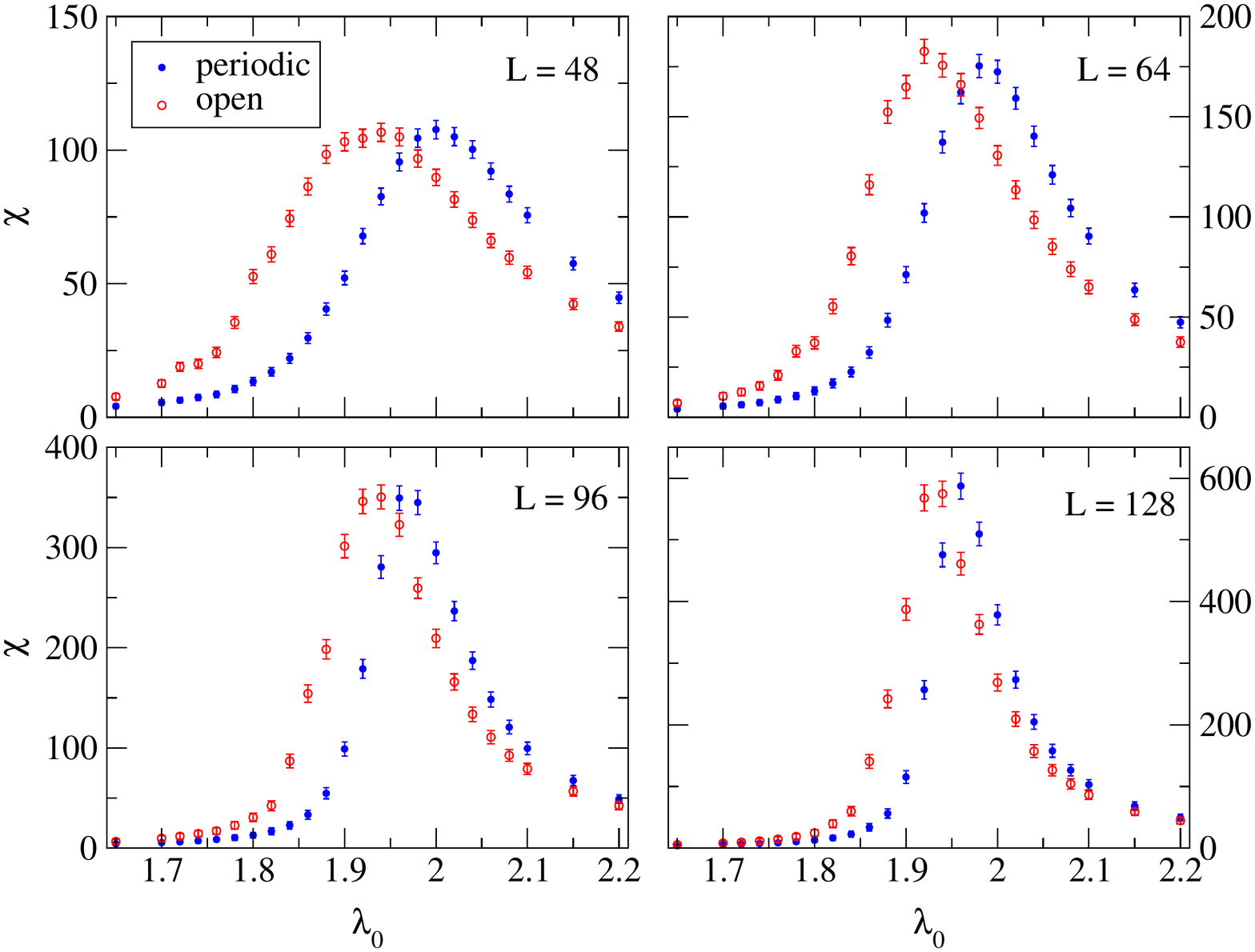}
\caption{Comparison of susceptibility between PBC and OPEN for different $L$
at $m_0^2=-0.5$ and gradient flow level 50.}
\label{susccompl}                                                               
\end{center}                                     
\end{figure}

Analogous to the study done for $\langle\mid\hspace{-0.2em}\phi\hspace{-0.2em}
\mid$, we compare the susceptibility between the periodic and open boundary
conditions for different lattice sizes (L) at $m_0^2=-0.5$ and gradient flow
level 50 in Fig. \ref{susccompl}. Here too, we observe that compared to the
case of PBC, the peak is shifted towards the smaller values of $\lambda_0$ in
case of OPEN at a fixed $L$ and the magnitude of the shift decreases as $L$
increases. 

\begin{figure}
\begin{center}   
\includegraphics[width=0.5\textwidth]{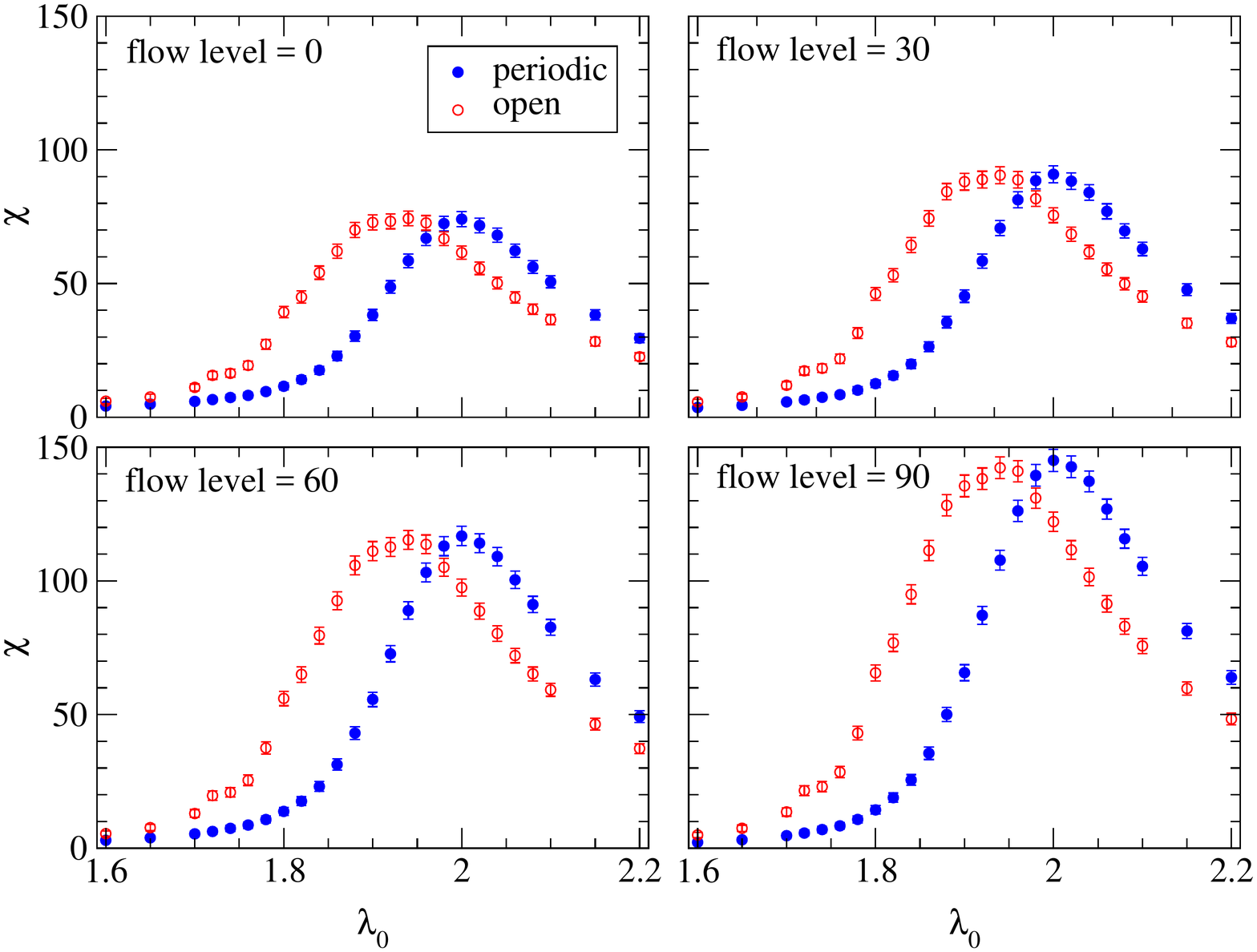}
\caption{Comparison of susceptibility between PBC and OPEN for different levels
of gradient flow at $m_0^2=-0.5$ with $L=48$.}
\label{susccompgfL48}
\end{center}                                     
\end{figure}

\begin{figure}
\begin{center}
\includegraphics[width=0.5\textwidth]{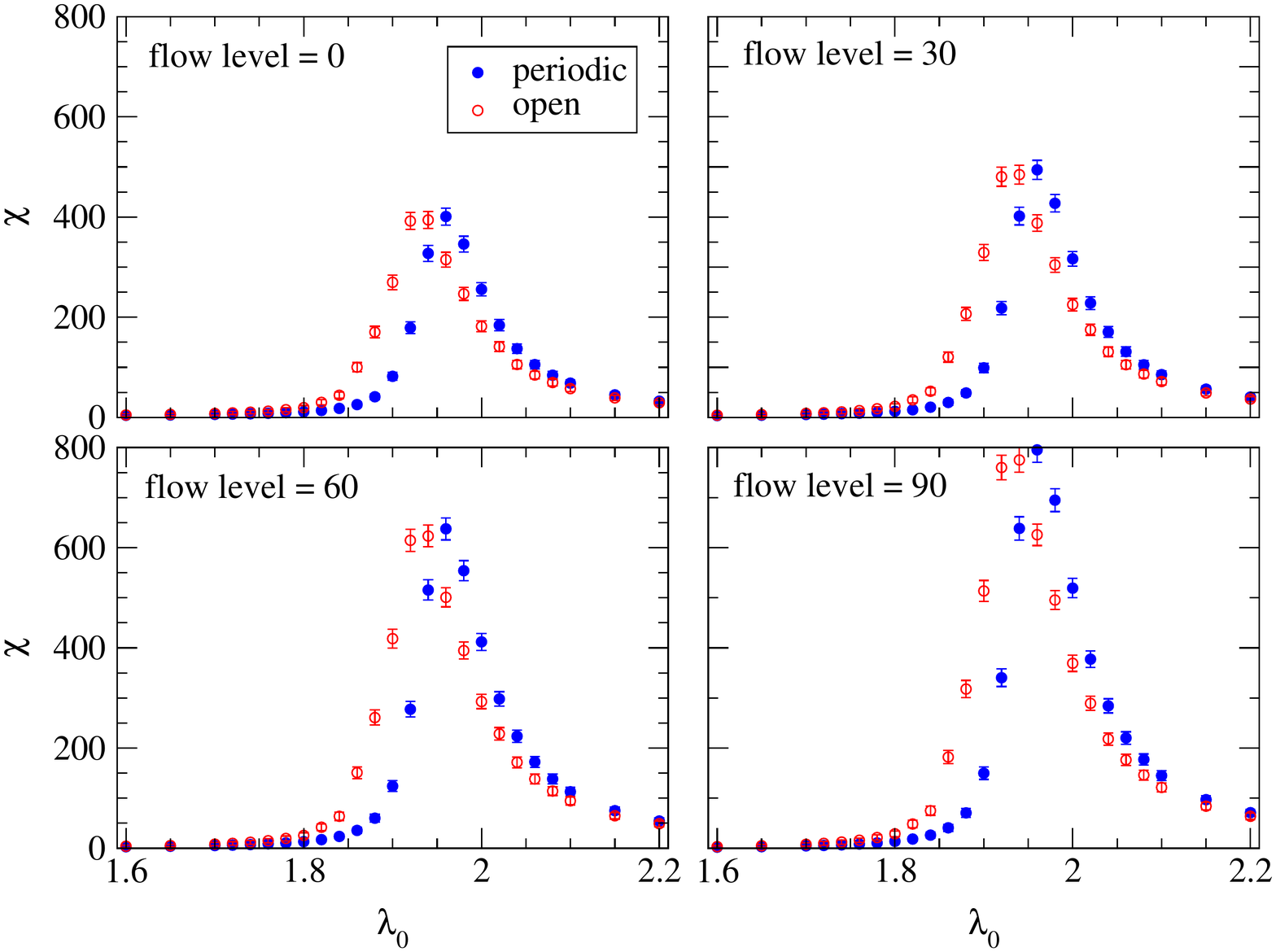}
\caption{Comparison of susceptibility between PBC and OPEN for different levels
of gradient flow at $m_0^2=-0.5$ with $L=128$.}
\label{susccompgfL128}                                              
\end{center}                                     
\end{figure}

This shift in the peak position of susceptibility between PBC and OPEN
seems to be unaffected by gradient flow as per our expectation. This has been
demonstrated in Figs. \ref{susccompgfL48} and \ref{susccompgfL128} for the
smallest and largest lattice sizes $L=48$ and $L=128$ respectively both with
$m_0^2=-0.5$.

From all the figures here comparing susceptibility between PBC and OPEN, it appears that the peak
is slightly sharper in the case of the former compared to the later consistent with the case of $\langle \mid \hspace{-0.2em} \phi \hspace{-0.2em} \mid
\rangle$.

\subsection{Finite size scaling analysis}
The fact that the value of $\langle \mid \hspace{-0.2em} \phi \hspace{-0.2em} \mid
\rangle$ changes with gradient flow even in the critical region raises an
interesting question. Does the determination of critical coupling in a Finite
Size Scaling (FSS) analysis gets affected by gradient flow? Here
we study the possible effect of gradient flow in the FSS of the data for
$\langle\mid\hspace{-0.2em}\phi\hspace{-0.2em}\mid\rangle$ to determine the
critical coupling. We follow the discussion of the main aspects  of finite size 
scaling \cite{cardy,brezin,goldenfeld} given in Ref. \cite{De:2005ny}. 
FSS assumes that, in a finite system, out of the three length scales involved,
namely, the correlation length $\xi$, the size of the system $L$ and the
microscopic length $a$ (lattice spacing), the last one drops out near the
critical region due to universality.

For any observable $P_L$ computed on a lattice of finite extent $L$, the
nonanaliticity near the critical point in the infinite volume limit can be
expressed in the form of a scaling law $P_{\infty}(\tau) = A_P\tau^{-\rho}$ where
$\tau$ = $(\lambda_0^c - \lambda_{0})/\lambda_0^c$ and $\rho$ is the critical
exponent associated with the observable. Following the arguments of FSS
analysis, it can be shown that 
\be L^{\rho/\nu}/P_L(\tau) = ~A_P^{-1}~A_{\xi}^{\rho/\nu}~\left[{\rm C_P}~ +
~ {\rm D_P}~A_{\xi}^{-1/\nu}~
\tau ~ L^{1/\nu}~+ ~ {\cal O}(\tau^2)\right] \label{fss2}
\ee
where $\nu$ is the critical exponent associated with the correlation length
$\xi$. Eq. (\ref{fss2}) implies that if we plot 
$L^{\rho/\nu}/P_L\left(\tau\right)$ versus the coupling $\lambda_0$ for different
values of $L$, all the curves will pass through the same point where $\tau=0$
or equivalently $ \lambda_0 = \lambda_0^c $ \cite{goldenfeld}. 

The critical behavior of $\langle\phi\rangle$, the susceptibility $\chi$ and
the mass gap $m=1/\xi$ may be written as 
\be
\langle\phi\rangle ~=~ A_{\phi}^{-1}~\tau^\beta, \quad
\chi ~=~ A_{\chi}~\tau^{-\gamma}\quad {\rm and}\quad
m ~=~ A_{\xi}^{-1}~\tau^\nu. \label{scaling}
\ee
From the general expectation that in $1+1$ dimensions, $\phi^4$ theory and Ising
model belong to the same universality class, we have used the Ising values for
the corresponding exponents as inputs in our FSS analysis. Thus, $\beta =
0.125$, $\gamma=1.75$ and $~\nu = 1$.
 
\begin{figure}
\begin{center}   
\includegraphics[width=0.5\textwidth]{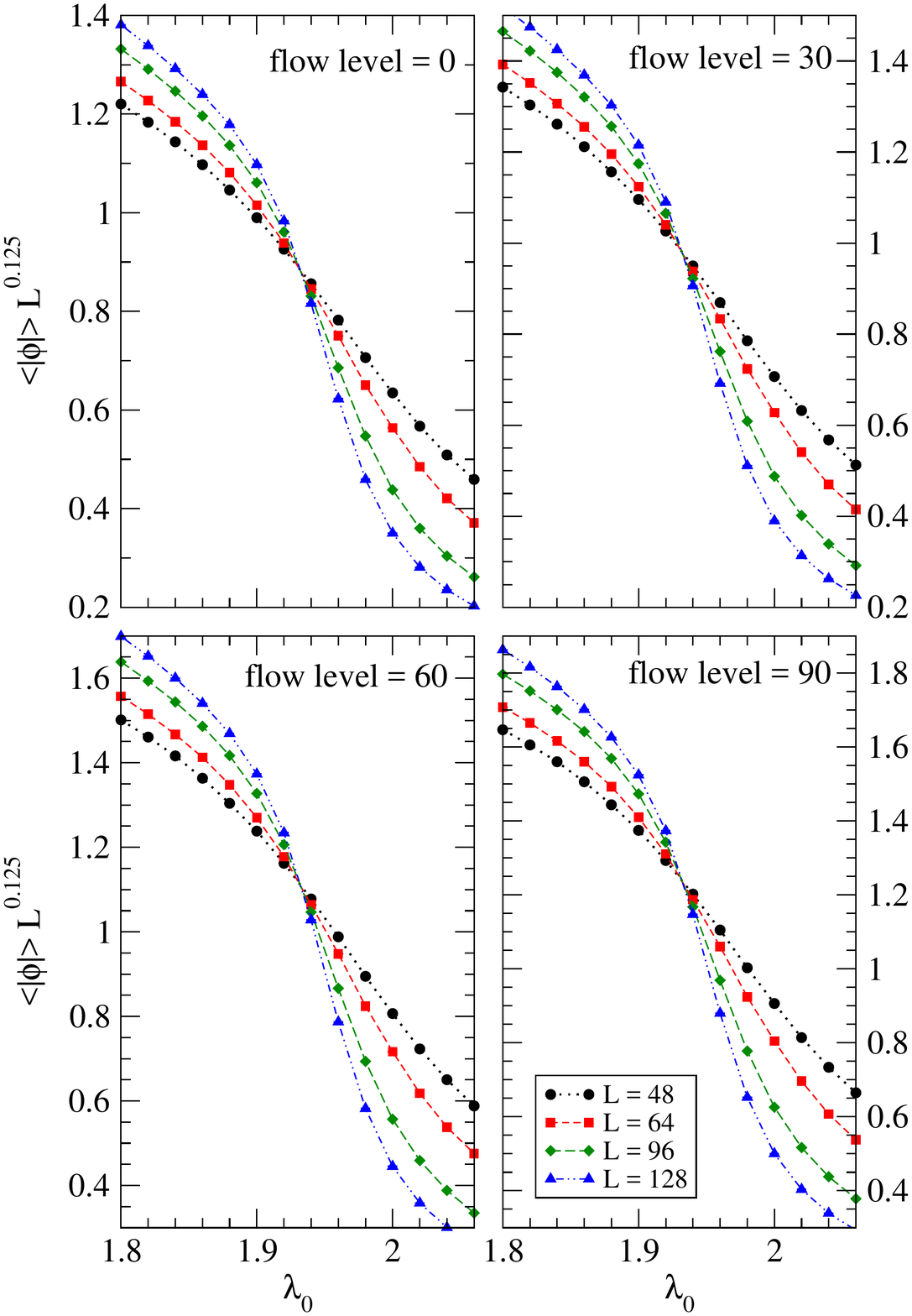}
\caption{Plot of $\langle \mid\hspace{-0.2em}\phi\hspace{-0.2em}\mid\rangle
L^{0.125}$ versus $\lambda_0$ for different levels of gradient flow with PBC at
$m_0^2=-0.5$.}
\label{fss-pbc}
\end{center}
\end{figure}

The plot of $\langle\mid\hspace{-0.2em}\phi\hspace{-0.2em}\mid\rangle L^{0.125}$
versus $\lambda_0$ for different values of the gradient flow level at $m_0^2 =
-0.5$ with periodic boundary in the temporal direction is presented in Fig.
\ref{fss-pbc}. In spite of the fact that $\langle\mid\hspace{-0.2em}\phi
\hspace{-0.2em}\mid\rangle$ changes with gradient flow level, the critical
coupling $\lambda_0^c$ is found to be unaffected by gradient flow. The critical
coupling $\lambda_0^c$ is found to be little less than $1.94$ for $m_0^2
= -0.5$. Similar FSS analysis has been done at $m_0^2 = -1.0$ leading to the
same conclusion along with an estimated value of $\lambda_0^c \approx 4.46$.  

\begin{figure}
\begin{center}   
\includegraphics[width=0.5\textwidth]{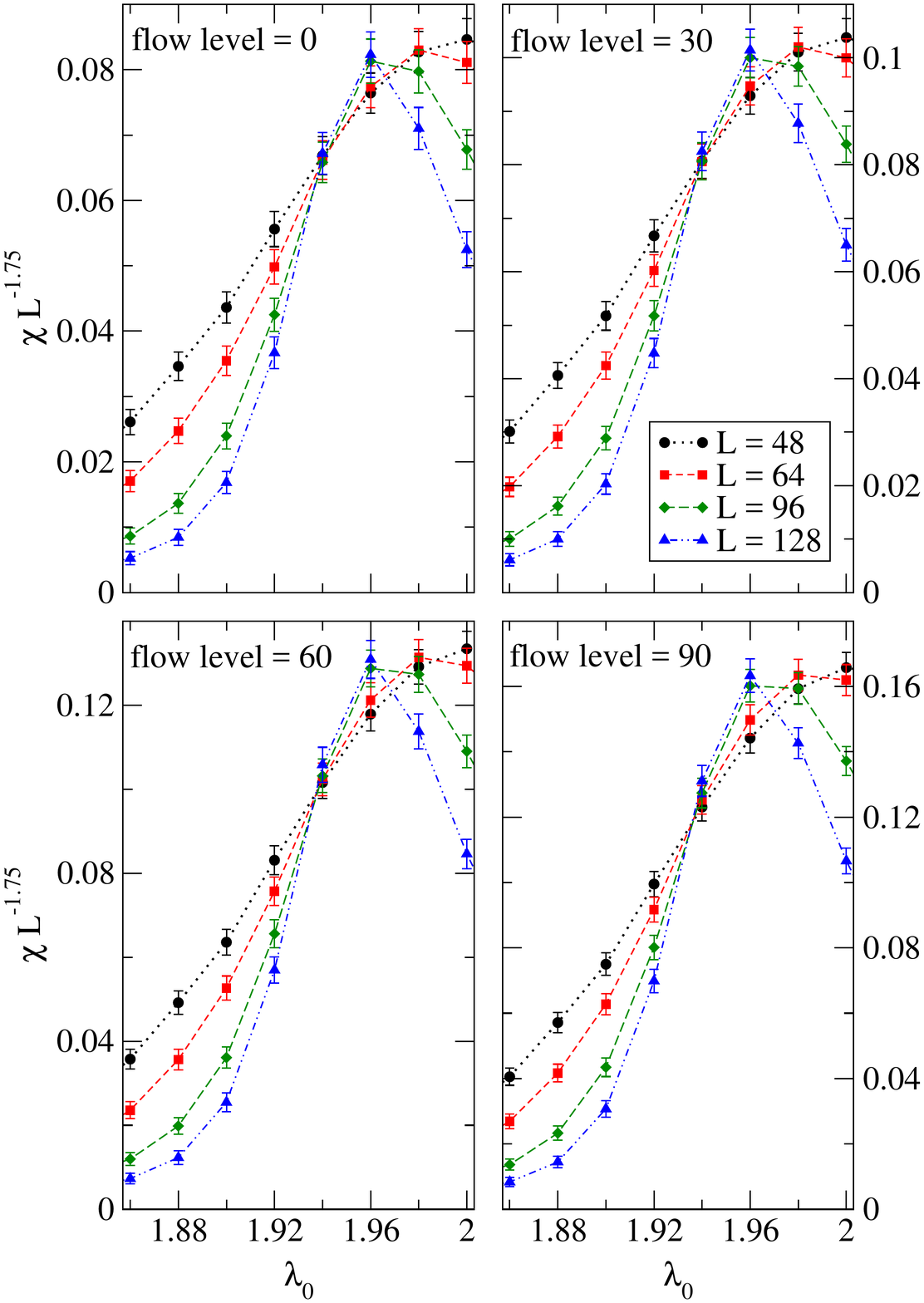}
\caption{Plot of $\chi L^{-1.75}$ versus $\lambda_0$ for different levels of
gradient flow with PBC at $m_0^2=-0.5$.}
\label{fss-chi-pbc}
\end{center}
\end{figure}

A similar FSS analysis of the susceptibility calculated at $m_0^2 = -0.5$ for
various values of gradient flow level (for PBC) is presented in
Fig. \ref{fss-chi-pbc}. Here, the smoothening effect of gradient flow has
helped to pinpoint the location of critical coupling $\lambda_0^c$ which matches
with the estimation done from the FSS analysis for $\langle \mid\hspace{-0.2em}
\phi\hspace{-0.2em}\mid\rangle$. Similar FSS analysis done at $m_0^2 = -1.0$
also gives the corresponding value of $\lambda_0^c$ matching with the same
obtained from FSS analysis for $\langle \mid\hspace{-0.2em}
\phi\hspace{-0.2em}\mid\rangle$.

\begin{figure}
\begin{center}   
\includegraphics[width=0.5\textwidth]{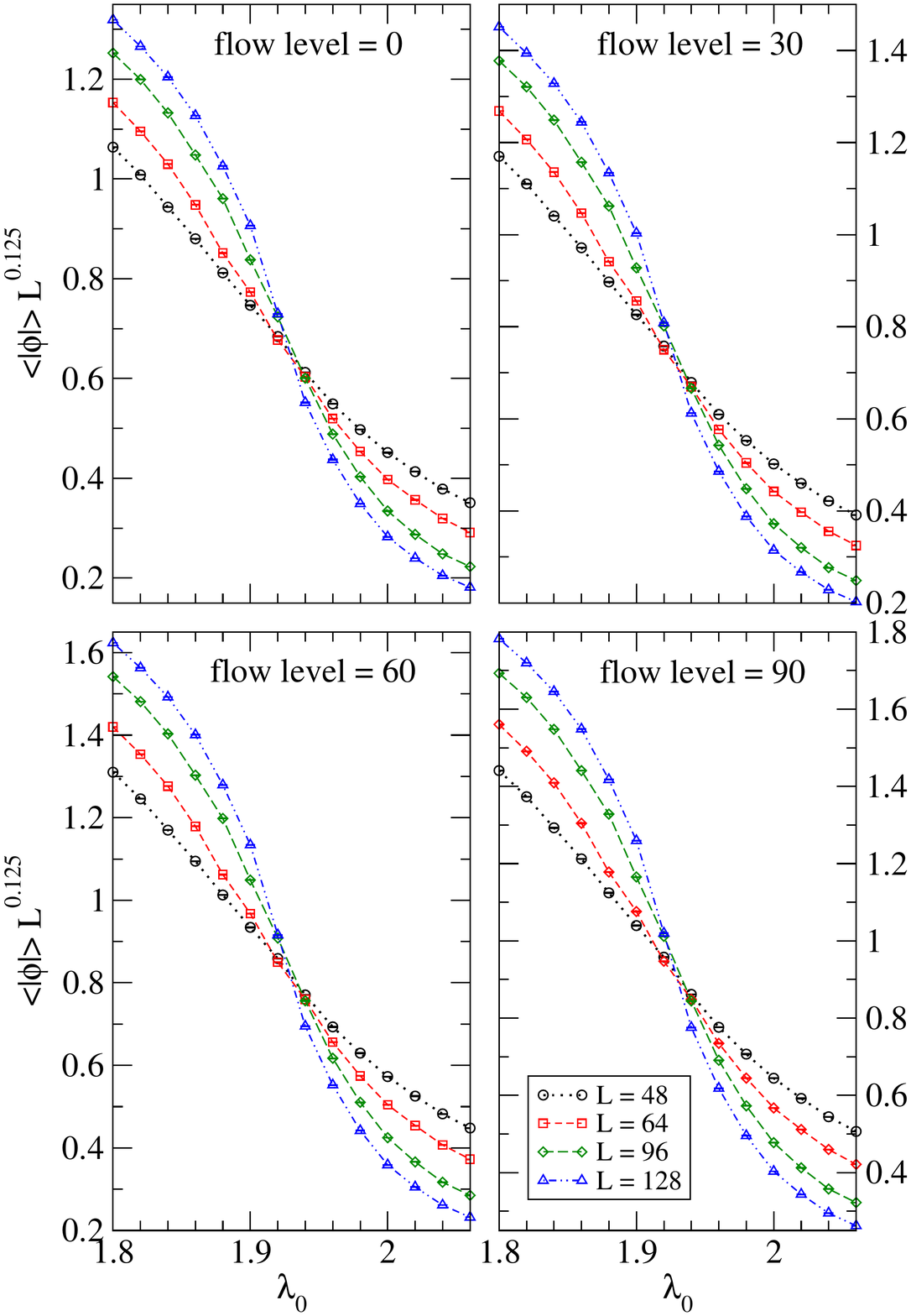}
\caption{Plot of $\langle \mid\hspace{-0.2em}\phi\hspace{-0.2em}\mid\rangle
L^{0.125}$ versus $\lambda_0$ for different levels of gradient flow at
$m_0^2=-0.5$ with open boundary.}
\label{fss-open}
\end{center}
\end{figure}

\begin{figure}
\begin{center}   
\includegraphics[width=0.5\textwidth]{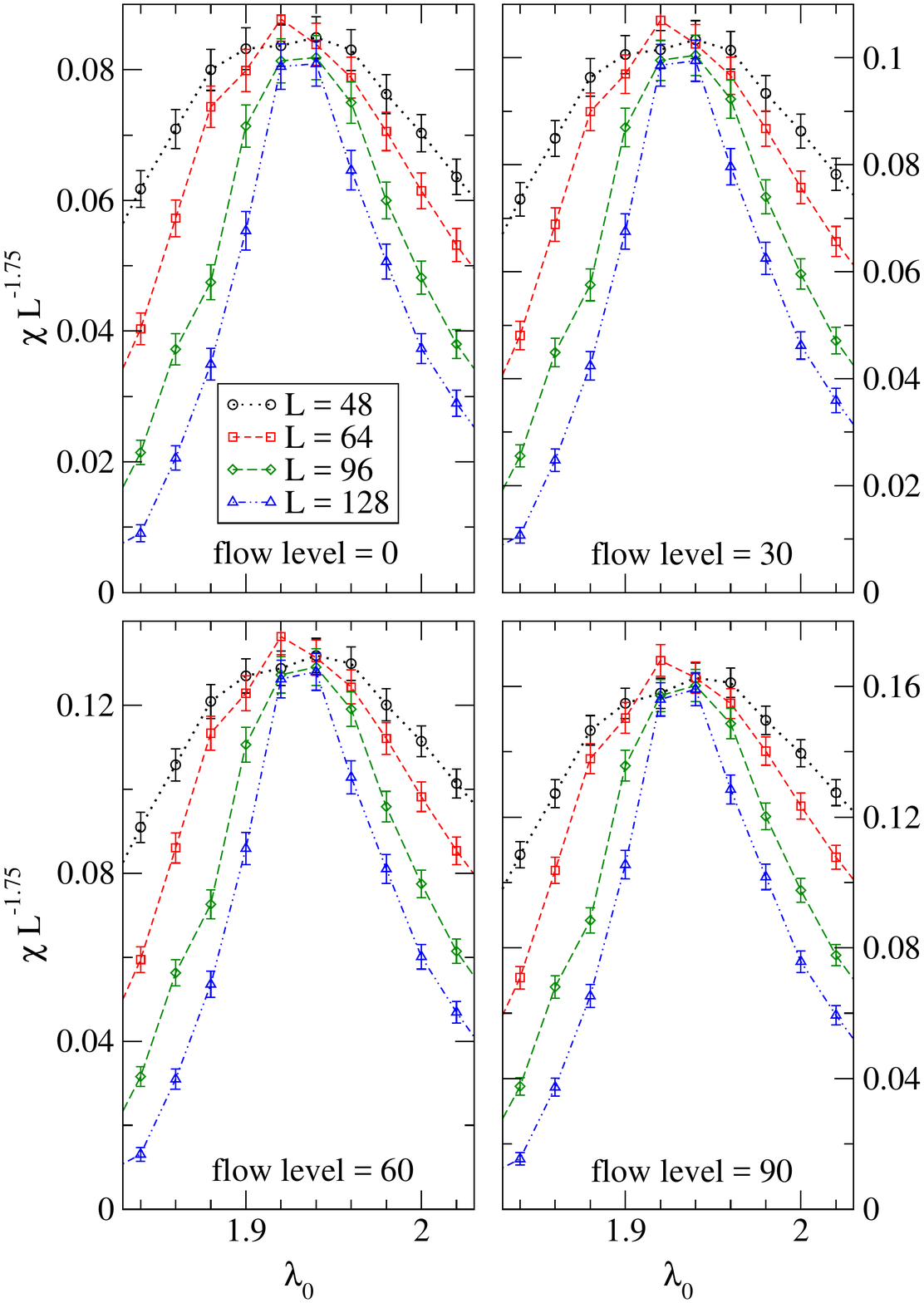}
\caption{Plot of $\chi L^{-1.75}$ versus $\lambda_0$ for different levels of
gradient flow at $m_0^2=-0.5$ with open boundary.}
\label{fss-chi-open}
\end{center}
\end{figure}

Using the same ansatz, we have tried an FSS analysis for both $\langle \mid
\hspace{-0.2em}\phi\hspace{-0.2em}\mid\rangle$ and $\chi$ in the case of open
boundary in the temporal direction. The results are respectively presented 
in Figs.
\ref{fss-open} and \ref{fss-chi-open}. It is obvious from the figures that
the critical point is not pinpointed that clearly compared to
 the case of PBC.
A strange pattern is found for susceptibility for which the curves with
different $L$, instead of passing through a point, appear to overlap within
statistical errors at and between the values $1.92$ and $1.94$ of $\lambda_0$.
Anyway, the effect of surface in the case of open boundary needs to be taken into
the ansatz for finite size scaling analysis. For that purpose a more precise
numerical study is necessary which is beyond the scope of this study.

\subsection{The boson mass}
Finally, in this subsection, we present the results for the mass 
spectrum of the theory. The boson
mass has been extracted from the plateau along the time direction for 
effective mass
which is computed from two point and one point correlation function of the time
sliced scalar field $\phi(t)$ respectively for PBC and OPEN. For this purpose,
we first find out the gradient flow level for which the plateau for effective
mass is found to be the most stable one separately for each coupling (also for
PBC and OPEN separately). The computation of mass has been done only
for the two largest lattices namely for $L=96$ and $L=128$ since, 
for the smaller
lattices, the temporal extent is not sufficiently long to get the fall of
correlation function particularly within the critical region. Because of this,
it is not meaningful to perform the formal FSS analysis. However, even with
lattices of only two different sizes, we observe noticeable finite volume
effect.

\begin{figure}   
\begin{center}   
\includegraphics[width=0.5\textwidth]{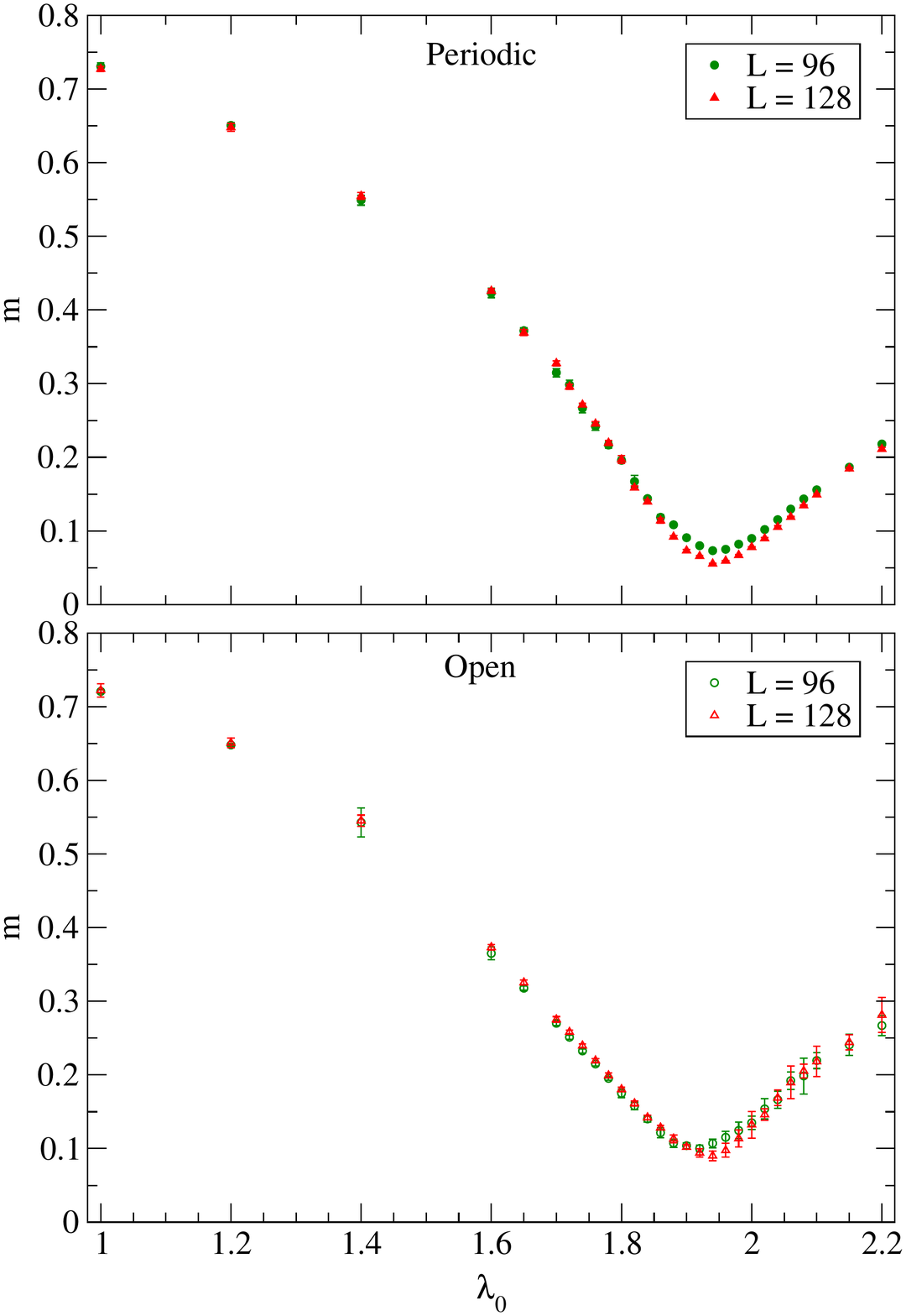}  
\caption{The volume dependence of the boson mass versus $\lambda_0$ for 
PBC (top) and OPEN (bottom) at $m_0^2=-0.5$.}
\label{massdiffl5}
\end{center}
\end{figure}

\begin{figure}
\begin{center}   
\includegraphics[width=0.5\textwidth]{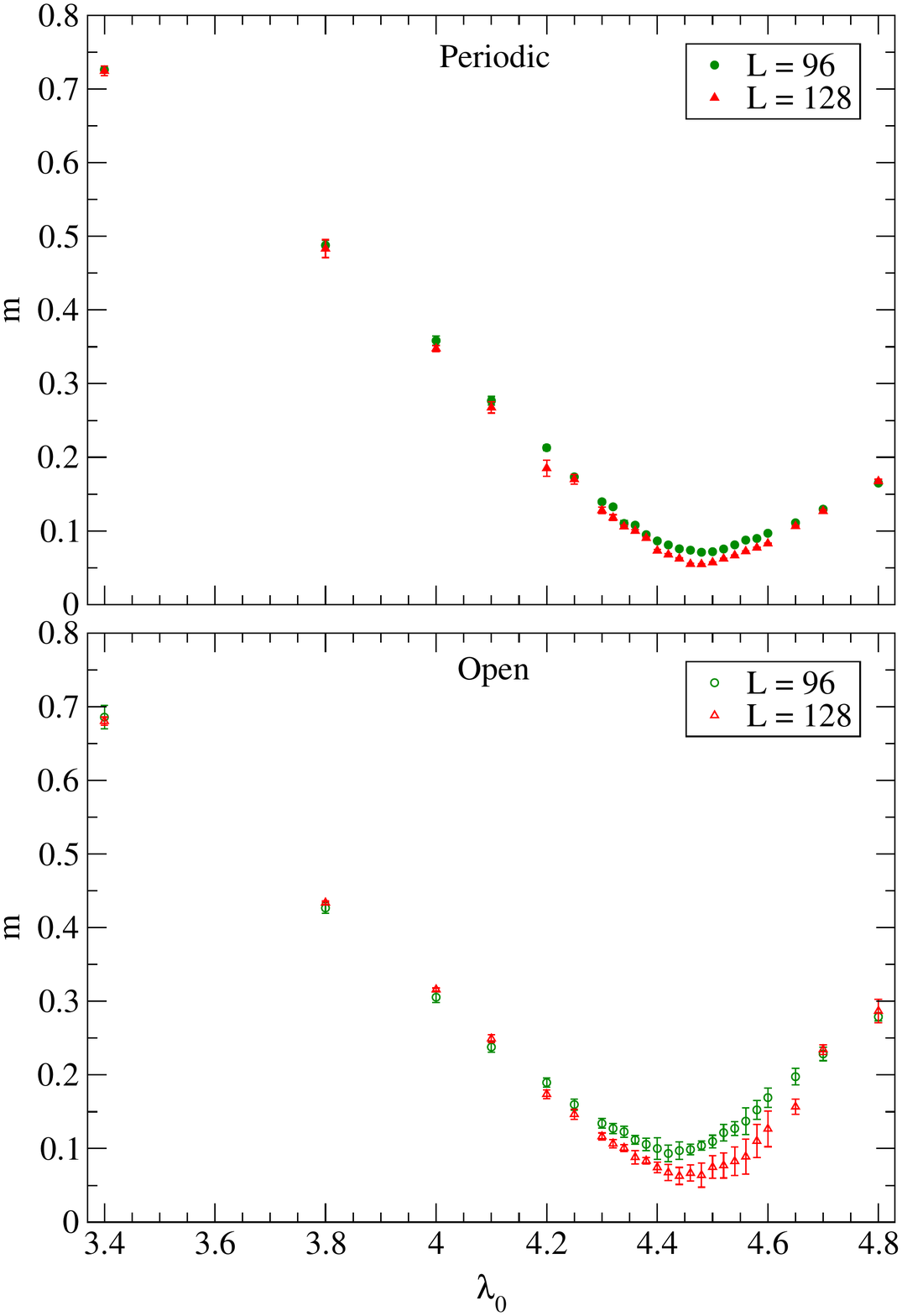}
\caption{The volume dependence of the boson mass versus $\lambda_0$ for 
PBC (top) and OPEN (bottom) at $m_0^2=-1.0$.}
\label{massdiffl1}                                                       
\end{center}
\end{figure}

In Figs. \ref{massdiffl5} and \ref{massdiffl1} we compare the volume
dependence of the boson mass versus $\lambda_0$ for periodic  
(top) and open (bottom) at $m_0^2=-0.5$ and  $m_0^2=-1.0$ respectively. 
In the case
of PBC, the effect of finite volume on the mass spectrum is clearly visible
in the critical region. Although for OPEN, this effect is not that obvious from
the respective figure. However, if we remember from the behaviors of $\langle
\mid\hspace{-0.2em}\phi\hspace{-0.2em}\mid\rangle$ and $\chi$ that the phase
transition point shifts towards the smaller values of $\lambda_0$ compared to
PBC and that the shift is more for smaller volume, it appears that this 
leftward shift of transition point has overshadowed the finite volume effect.


\begin{figure}   
\begin{center}   
\includegraphics[width=0.5\textwidth]{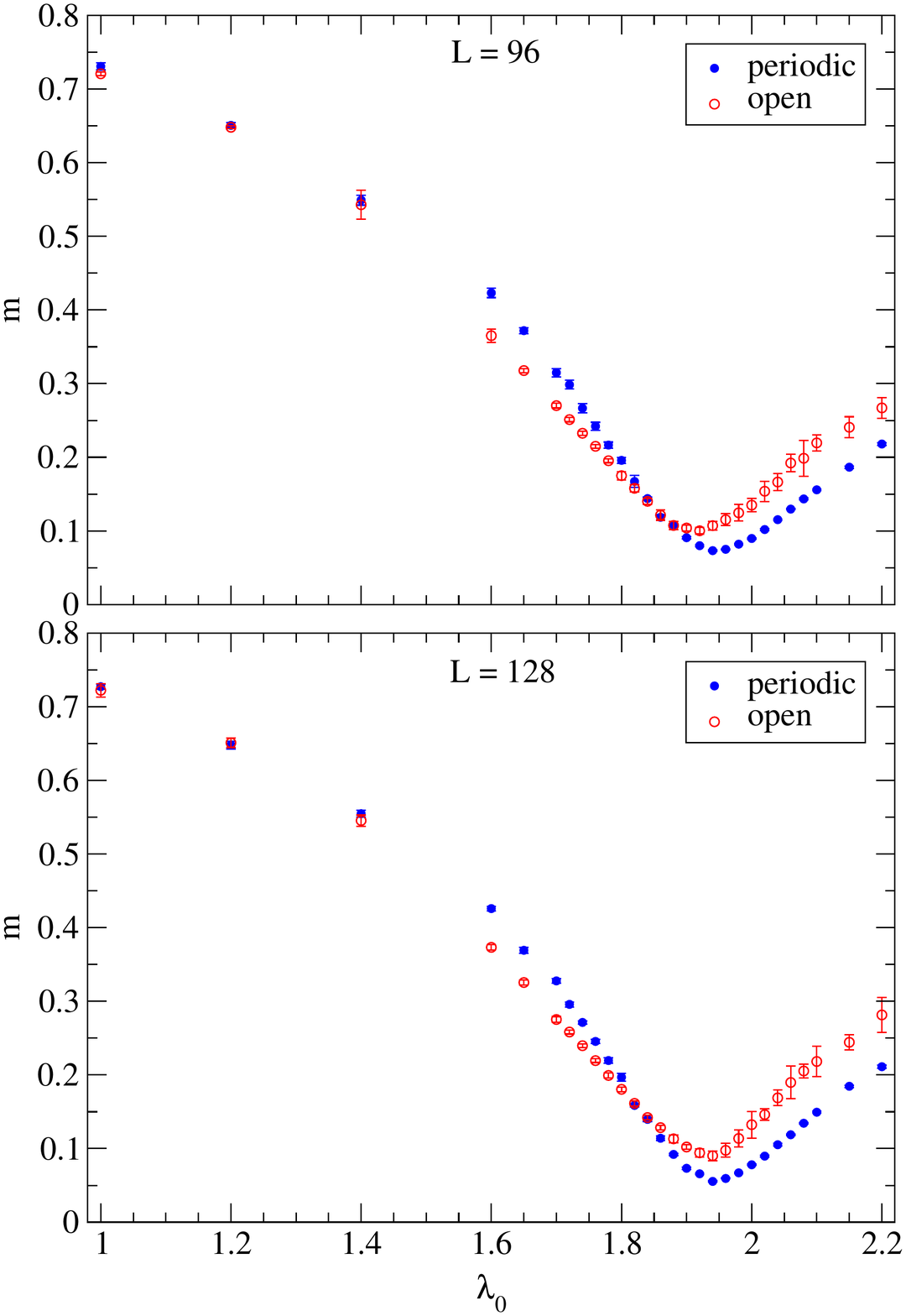}  
\caption{The effect of boundary condition on the boson mass versus 
$\lambda_0$ for L=96 (top) and L=128 (bottom) at $m_0^2=-0.5$.}
\label{masspo5}
\end{center}
\end{figure}

\begin{figure}
\begin{center}   
\includegraphics[width=0.5\textwidth]{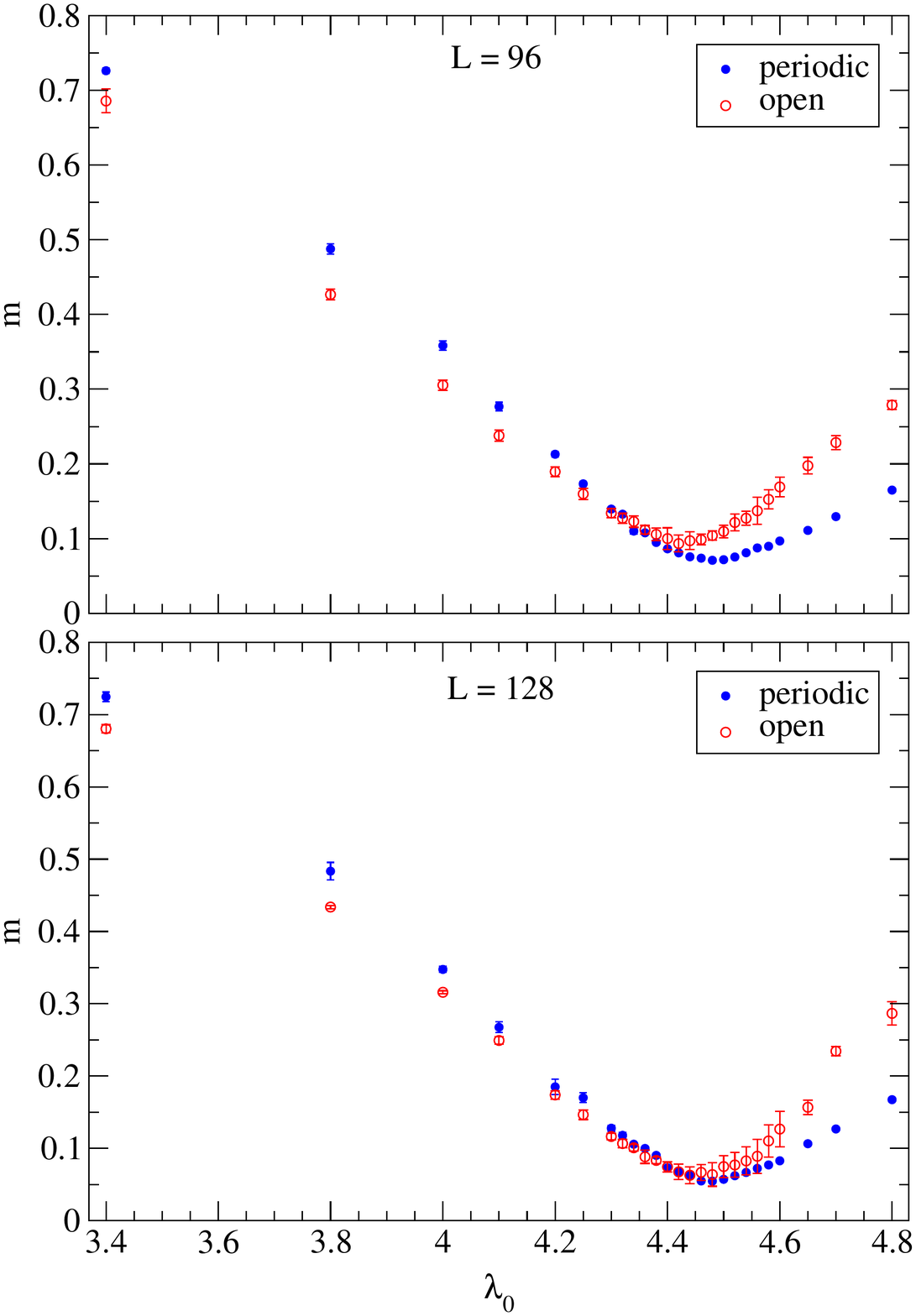}
\caption{The effect of boundary condition on the boson mass versus
$\lambda_0$ for L=96 (top) and L=128 (bottom) at $m_0^2=-1.0$.}
\label{masspo1}                                                       
\end{center}
\end{figure}

The effects of the boundary on the mass spectrum have been demonstrated
in Figs. \ref{masspo5} and \ref{masspo1} respectively for two different
values for the lattice mass parameters $m_0^2=-0.5$ and $m_0^2=-1.0$ 
respectively. The leftward shift of the phase transition point is found to be almost
diminishing for the larger lattice. The boundary effects are more prominent
near the critical point as evident from the clearly sharper nature of transition in the case of PBC
compared to that of OPEN.

\section{Conclusions}
We have found the gradient flow to be very effective in reducing the 
unwanted lattice artifacts in the functional average of the time 
sliced field $\phi(t)$, the susceptibility and in the extraction of boson mass
from both two-point (PBC) and one-point (OPEN) correlators. We have shown that,
in spite of the fact that $\langle\mid\hspace{-0.2em}\phi\hspace{-0.2em}\mid\rangle$
changes with gradient flow (transition more prominent with increasing flow level),
the transition point does not depend on the gradient flow level. In the case of
susceptibility, it is found that the increase in the level of gradient flow raises
the height of the peak but leaves the peak position (phase transition point for a
given volume) practically unaltered. This behaviour is consistent with the expectation
that the gradient flow helps to reduce the lattice artifacts and brings the lattice
theory closer to the continuum physics where we expect the susceptibility to diverge
as critical point is approached. Critical coupling $\lambda_0^c$ has been obtained for
both $\langle\mid\hspace{-0.2em}\phi\hspace{-0.2em}\mid\rangle$ and $\chi$ from a
detailed finite size scaling analysis of the data for various lattice sizes and
gradient flow levels with results qualitatively unchanged with flow level but seem to
be pin-pointed more clearly with nonzero value of flow level.

With all kinds of observables studied here, for a given volume, phase transition point
has a left-ward shift in terms of $\lambda_0$ in the case of OPEN compared to 
the PBC for a
fixed value of $m_0^2$ and this left-ward shift diminishes as volume increases.
In addition, for a given volume, the phase transition curve is found to be more
prominent or sharp in the case of PBC compared to the OPEN. Particularly, noticeable boundary
effects are observed on the mass spectrum in the critical region where the effects of open boundary extend deeply into the bulk region (as emphasized by the behavior of
$\langle\mid\hspace{-0.2em}\phi(t)\hspace{-0.2em}\mid\rangle$) and the finite volume
artifacts become much more important compared to the case of the periodic boundary condition.
An extensive analytical study of the boundary effects due to open boundary in the critical
region, taking into consideration the finite size scaling will be very fruitful.      

The main objective of the present study has been to investigate
 the effects of the gradient flow and the open boundary condition in the
temporal direction in a theory with vanishing mass gap and without the
complexities of renormalization. Since it is known that in 3+1 dimensional
$\phi^4$ theory, the straight forward use of the gradient flow equation
leads to new divergences in correlation functions at non-zero flow time, 
a detailed comparison of various proposals to overcome this problem needs to
be investigated. Our current study has demonstrated that in the region of vanishing
mass gap, open boundary introduces complexities when used in a lattice with
finite volume. Our previous success with open boundary in pure Yang-Mills 
theory may be partly due to the reasonably large mass gap (glueball mass) in
this theory. On the other hand the relevant mass gap of QCD (determined by 
the two pion state) is much lower and we expect open boundary to have
non-trivial consequences in the scaling region. This requires a thorough
investigation in the future.        

\begin{acknowledgments}
To carry out all the numerical calculations reported in this work, Cray XE6 
system supported by the 11th-12th Five Year
Plan Projects of the Theory Division, SINP under the Department of
Atomic Energy, Govt. of India, is used.
We thank Richard Chang for the prompt maintenance of the system.
\end{acknowledgments}

\end{document}